\documentclass[a4paper,10pt]{article}
\usepackage{a4wide}
\usepackage{graphicx}
\usepackage{amssymb}
\usepackage{amsmath}
\usepackage{amsthm}
\usepackage{url}

\newtheorem{definition}{Definition}

\newcommand{\df}{\ensuremath{\mathrel{:=}}}

\newcommand{\Union}{\ensuremath\bigcup}

\newcommand{\logAND}{\bigwedge}

\newcommand{\natnum}{\mathbb{N}}    

\newcommand{\fni}[1]{{\text{\sf #1}}}  

\newcommand{\Schema}{\mathcal{S}}
\newcommand{\DB}{\mathcal{D}}

\newcommand{\Attr}{\mathcal{A}}
\newcommand{\Tables}{\mathcal{T}}

\newcommand{\Proj}{\mathit{Proj}}
\newcommand{\Sel}{\mathit{Sel}}

\newcommand{\id}{\mathit{id}}

\newcommand{\TS}{\mathit{TS}}
\newcommand{\Inst}{\mathit{Inst}}

\newcommand{\ie}{i.e.,\ }

\newcommand{\st}{s.t.\ }

\newcommand{\Path}{\mathit{Path}}
\begin{document}
%
%
\title{Artifact Lifecycle Discovery}
\author{Viara Popova\textsuperscript{1}, Dirk Fahland\textsuperscript{2}, Marlon Dumas\textsuperscript{1} \\
\textsuperscript{1} Institute of Computer Science, \\ University of Tartu, J. Liivi 2, \\Tartu 50409, Estonia \\
\texttt{\{viara.popova, marlon.dumas\}@ut.ee}\\
\textsuperscript{2} Eindhoven University of Technology \\ The Netherlands \\
\texttt{d.fahland@tue.nl}
}

\maketitle

\begin{abstract}        
\noindent Artifact-centric modeling is a promising approach for modeling business processes based on
the so-called business artifacts - key entities driving the company's operations and whose lifecycles
define the overall business process. While artifact-centric modeling shows significant advantages,
the overwhelming majority of existing process mining methods cannot be applied (directly) 
as they are tailored to discover monolithic process models.
This paper addresses the problem by proposing a chain of methods that can be applied to discover 
artifact lifecycle models in Guard-Stage-Milestone notation.
We decompose the problem in such a way that a wide
range of existing (non-artifact-centric) process discovery and analysis methods can be reused in a flexible manner.
The methods presented in this paper are implemented as software plug-ins for ProM, a generic open-source framework
and architecture for implementing process mining tools.\\
                 
\noindent {\bf Keywords:} Artifact-Centric Modeling, Process Mining, Business Process Modeling
\end{abstract}
\section{Introduction}
Traditional business process modeling is centered around the process and other aspects such as
data flow remain implicit, buried in the flow of activities. 
However, for a large number of processes, the flow of activities is inherently intertwined 
with the process' data flow, often to the extent that a pure control-flow model cannot capture 
the process dynamic correctly. A prime example is a build-to-order process where several customer 
orders are collected, and based on the ordered goods multiple material orders are created. 
Typically, one customer order leads to several material orders and one material order contains 
items from several different customer orders. These n-to-m relations between customer orders 
and material orders influence process dynamics, for instance if a customer cancels her order. 
Such complex dynamics cannot be represented in a classical activity-flow centric process model.

Artifact-centric modeling is a promising approach for modeling business processes based on
the so-called business artifacts~\cite{artifacts2, artifacts1} -- key entities driving the company's operations and whose lifecycles
define the overall business process. An artifact type contains an information model with all
data relevant for the entities of that type as well as a lifecycle model which specifies how
the entity can progress responding to events and undergoing transformations from its creation until it is archived. 

Most existing work on business artifacts has focused on the use of lifecycle models based on variants
of finite state machines. Recently, a new approach was introduced -- the Guard-Stage-Milestone (GSM)
meta-model~\cite{gsm2, gsm1} for artifact lifecycles. GSM is more declarative
than the finite state machine variants, and supports hierarchy
and parallelism within a single artifact instance. 

Some of the advantages of GSM~\cite{gsm2, gsm1} are in the intuitive nature of the used constructs which reflect the way
stakeholders think about their business. Furthermore, its hierarchical structure allows for 
a high-level, abstract view on the operations while still being executable. It supports a wide range of 
process types, from the highly prescriptive to the highly descriptive. It also provides
a natural, modular structuring for specifying the
overall behavior and constraints of a model of business operations
in terms of ECA-like rules.

Process mining includes techniques for discovery and analysis of business process models (such as
conformance checking, repair, performance analysis, social networking and so on) from event logs
describing actual executions of the process.
While artifact-centric modeling in general and GSM in particular show significant advantages,
the overwhelming majority of existing process mining methods cannot be applied (directly) in such a setting.
The prime reason is that existing process mining techniques are tailored to classical monolithic processes 
where each process execution can be described just by the flow of activities. 
When applied to processes over objects in n-to-m relations (expressible in artifacts), 
classical process mining techniques yield incomprehensible results due to numerous side effects~\cite{FahlandLDA11}.

This paper addresses the problem by proposing a chain of methods that can be applied to discover 
artifact lifecycle models in GSM notation. We decompose the problem in such a way that a wide
range of existing (non-artifact-centric) process discovery and analysis methods can be reused 
in the artifact-centric setting in a flexible manner. Our approach is described briefly in the
following paragraphs.

Typically, a system that executes a business process in a process-centric setting 
records all events of one execution in an isolated case; all
cases together form a log. The cases are isolated from each other:
each event occurs in exactly one case, all events of the case together
describe how the execution evolved. 

Traditional methods for automatic process discovery assume that different process executions are recorded in separate cases. 
This is not suitable in the context of artifact-centric systems.
Indeed, artifact-centric systems allow high level
of parallelism and complex relationships between multiple instances of artifacts~\cite{FahlandLDA11}. 
This can result in overlapping cases and one
case can record multiple instances of multiple artifacts. 
Therefore, we  do not assume that the logs given as input are structured in terms of cases.
Instead, all events may be recorded together, without any case-based grouping, as a
raw event log. 

Such a raw log contains all observed events where each event includes a timestamp reflecting when the
event occurred and a collection of attribute-value pairs, 
representing data that is read, written or deleted due to the occurrence
of the event.
Fig.~\ref{fig:RawLogExample} gives an example of how such a raw log might
look like. The specific format might differ depending on the system that generates it.

Taking such a log as a starting point, we propose a tool chain of methods that can produce a model
in GSM notation which reflects the behavior demonstrated in the event log. We decompose the problem in three main steps
which would allow us to take advantage of the vast amount of existing research in process mining
and reuse some of the existing tools for process discovery. 

\begin{figure}[t]\centering
    \includegraphics[width=0.8\linewidth]{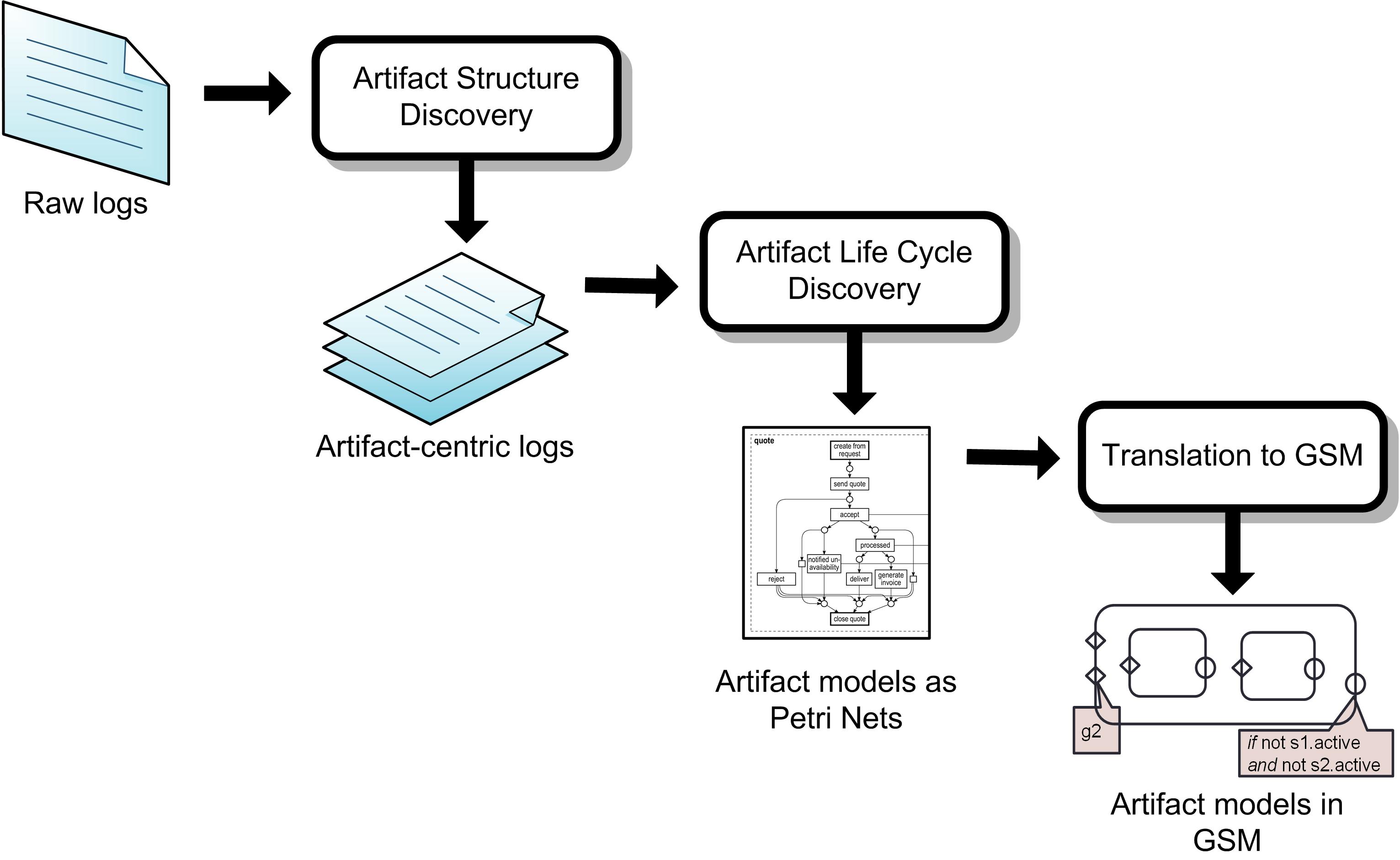}
    \caption{The overall architecture of the artifact lifecycle discovery process.}
    \label{fig:architecture}
\end{figure}

Fig.~\ref{fig:architecture} shows the proposed
overall architecture of the artifact lifecycle discovery process. 
First, based on the data incorporated in the events,
the artifact decomposition of the problem is determined, i.e. how many artifacts can
be found and which events belong to which instances of which artifact. Using this 
information, the log can be decomposed to generate so-called artifact-centric logs where
each trace contains the event for one instance of one artifact and each log groups
the traces for one artifact. 

The artifact-centric logs can then be used to discover the lifecycle of each
artifact. For this, any existing method can be used most of which generate models in Petri Net notation. 
Therefore, as a final step, we apply a method for translating Petri Net models into
GSM notation. 

The methods presented in this paper are implemented as software plug-ins for ProM~\cite{V_B_vD_vdA@BPM2010}, a generic open-source framework
and architecture for implementing process mining tools. The implementation is part of the \fni{ArtifactModeling}
package which is available from \fni{www.processmining.org}.

The paper is organized as follows. Section~\ref{sec:build-to-order} presents a working example used for illustration
of the proposed methods and Section~\ref{sec:background} reviews the necessary background knowledge. 
Section~\ref{sec:art-struct} presents the artifact structure discovery step of the tool chain. 
Section~\ref{sec:lifecycle-disc} discusses the artifact lifecycle discovery step and Section~\ref{sec:PN2GSMgen} presents the 
method for translating Petri net models to GSM. Finally, Section~\ref{sec:conclusions} concludes the paper
with a discussion and directions for future work.
\section{The Build-to-Order Scenario}
\label{sec:build-to-order}
As a motivating example used to illustrate the proposed methods, we consider a build-to-order process as follows.
The process starts when the manufacturer receives a purchase order from a customer
for a product that needs to be manufactured. 
This product typically requires multiple components or materials
which need to be sourced from suppliers. To keep track of this process, the manufacturer
first creates a so-called work order which includes multiple line items --
one for each required component. Multiple suppliers can supply the same materials
thus the manufacturer needs to select suppliers first, then place a number of
material orders to the selected ones. 

Suppliers can accept or reject the orders.
If an order is rejected by the supplier then a new supplier is found for these components.
If accepted, the items are delivered and, in parallel, an invoice is sent to
the manufacturer. If the items are of sufficient quality then they are assembled into the product.
For simplicity we do not include here the process of returning the items, renegotiating with
the supplier and so on. Instead it is assumed that the material order will be marked as failed
and a new material order will be created for the items to a different supplier. 
When all material orders for the same purchase order are received and assembled, the product 
is delivered to the customer and an invoice is sent for it.

\begin{figure}[h]
\centering
\includegraphics[height=0.3\textwidth]{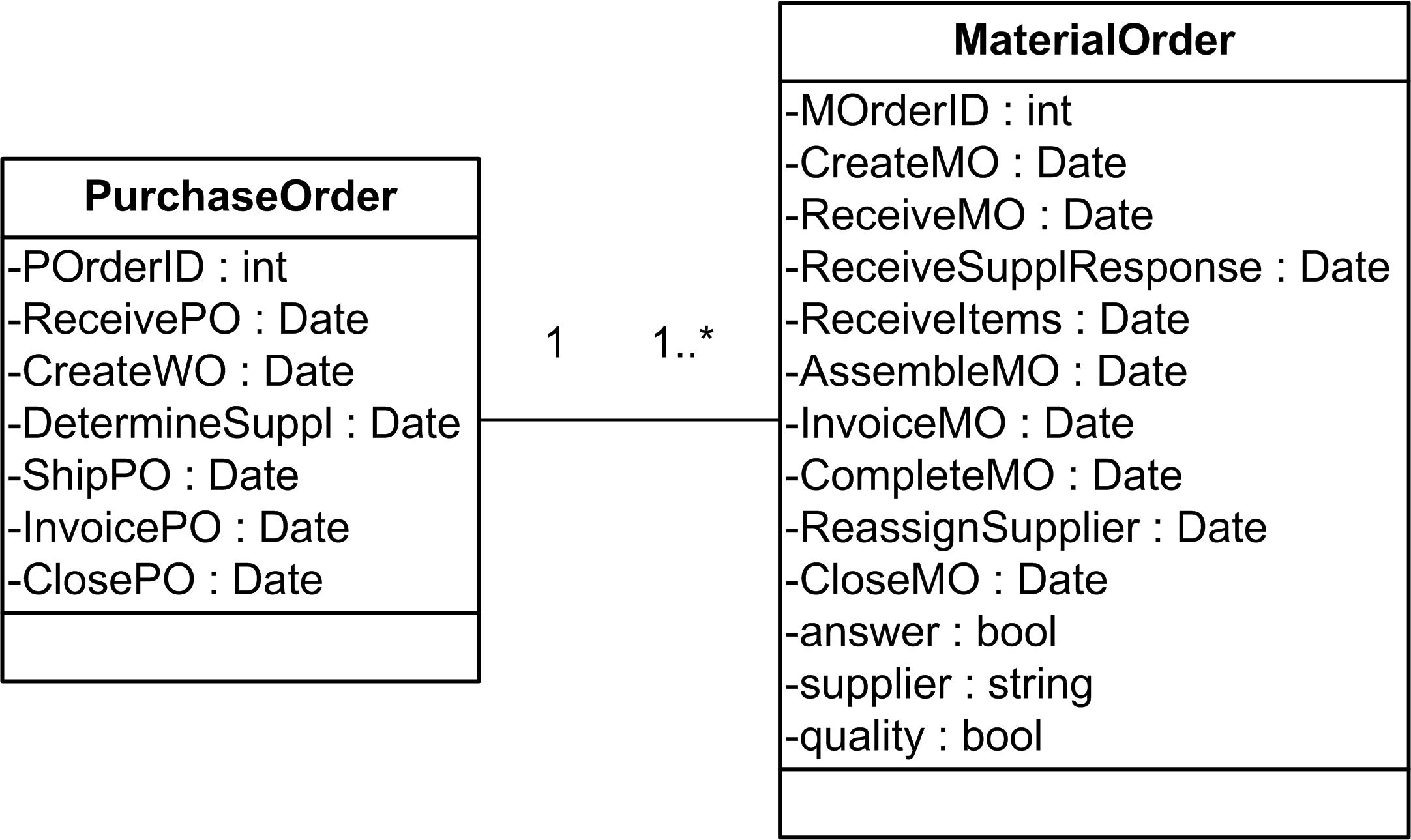}
\caption{The underlying data model of the build-to-order process.}
\label{fig:ER-model}
\end{figure}

Fig.~\ref{fig:ER-model} shows the underlying data model for the build-to-order example
which indicates that we can distinguish two artifact types: Purchase order and Material order
and one Purchase order corresponds to one or more Material orders. 

\begin{figure}[t]\centering
\includegraphics[width=.24\textheight]{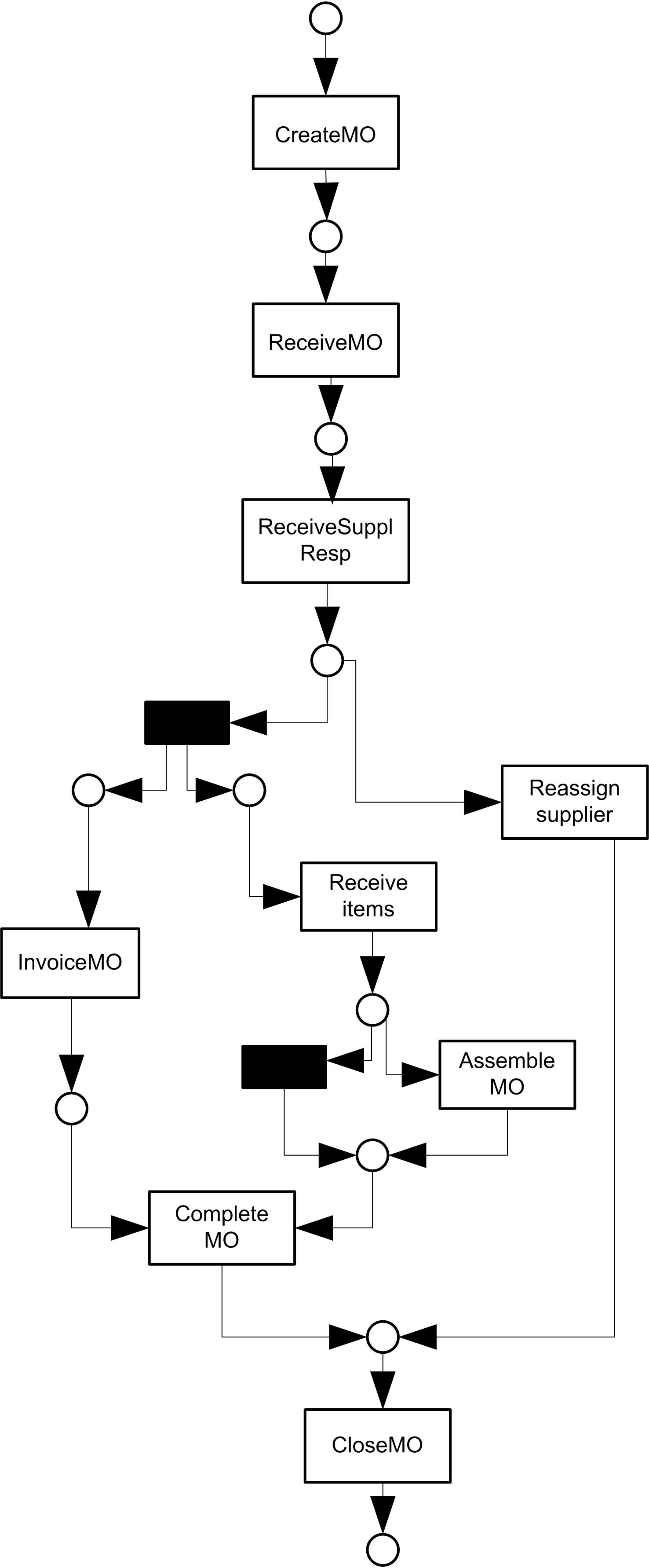}
\caption{A Petri net model for the lifecycle of the Material Order artifact, part of the build-to-order example}
\label{fig:ordertocash}
\end{figure}

Fig.~\ref{fig:ordertocash} shows one way of modeling the lifecycle of the Material order
artifact type using Petri net notation.
The details of the Petri net notation are explained in the next section.
Fig.~\ref{fig:RawLogExample} gives an example of how a raw log recording actual executions of 
a build-to-order process might look like.
\begin{figure}\centering{\sf\small
\begin{tabular}{lll}
11-24,17:12  & ReceivePO  & items=(it0), POrderID=1\\
11-24,17:13  & CreateMO   & supplier=supp6, items=(it0), POrderID=1,  MOrderID=1\\
11-24,19:56  & ReceiveMO  & supplier=supp6, items=(it0), POrderID=1, MOrderID=1\\
11-24,19:57  & ReceiveSupplResp    & supplier=supp6, items=(it0), POrderID=1, MOrderID=1, answer=accept\\
11-25,07:20  & ReceiveItems       & supplier=supp6, items=(it0), POrderID=1, MOrderID=1 \\
11-25,08:31 & Assemble & items=(it0), POrderID=1, MOrderID=1\\
11-25,08:53 & ReceivePO    & items=(it0,it1,it2,it3), POrderID=2\\
11-25,12:11 & ShipPO & POrderID=1\\
11-26,09:30 & InvoicePO & POrderID=1\\
11-26,09:31 & CreateMO & supplier=supp1, items=(it1,it2,it3), POrderID=2, MOrderID=2\\
11-28,08:12 & ReceiveMO & supplier=supp1, items=(it1,it2,it3), POrderID=2, MOrderID=2\\
11-28,12:22 & CreateMO & supplier=supp4, items=(it0), POrderID=2, MOrderID=3\\
12-03,14:34 & ClosePO & POrderID=1\\
12-03,14:54 & ReceiveMO & supplier=supp4, items=(it0), POrderID=2, MOrderID=3\\
12-03,14:55 & ReceiveSupplResp & supplier=supp1, items=(it1,it2,it3), POrderID=2, MOrderID=2, answer=accept\\
12-04,15:02 & ReceivePO  & items=(it1,it2), POrderID=3\\
12-04,15:20 & ReceiveSupplResp    & supplier=supp4, items=(it0), POrderID=2, MOrderID=3, answer=accept\\
12-04,15:33 & CreateMO   & supplier=supp2, items=(it2), POrderID=3, MOrderID=4\\
12-04,15:56 & ReceiveMO   & supplier=supp2, items=(it2), POrderID=3, MOrderID=4\\
12-04,16:30 & CreateMO   & supplier=supp5, items=(it1), POrderID=3, MOrderID=5\\
12-05,09:32 & ReceiveMO   & supplier=supp5, items=(it1), POrderID=3, MOrderID=5\\
12-05,09:34 & ReceiveItems       & supplier=supp4, items=(it0), POrderID=2, MOrderID=3\\
12-05,11:33 & ReceiveItems       & supplier=supp1, items=(it1,it2,it3), POrderID=2, MOrderID=2\\
12-05,11:37 & Assemble & items=(it0), POrderID=2, MOrderID=3\\
12-05,11:50 & ReceiveSupplResp    & supplier=supp2, items=(it2), POrderID=3, MOrderID=4, answer=reject\\
12-05,13:03 & ReceiveSupplResp    & supplier=supp5, items=(it1), POrderID=3, MOrderID=5, answer=accept\\
12-06,05:23 & Assemble & items=(it1,it2,it3), POrderID=2, MOrderID=2\\
12-06,05:25 & ReassignSupplier & items=(it2), POrderID=3, MOrderID=4\\
12-06,07:14 & ReceiveItems       & supplier=supp5, items=(it1), POrderID=3, MOrderID=5\\
12-06,07:15 & Assemble & items=(it1), POrderID=3, MOrderID=5\\
12-06,07:25 & InvoicePO & POrderID=2\\
12-06,09:34 & ShipPO & POrderID=2\\
12-12,20:41 & CreateMO   & supplier=supp5, items=(it2), POrderID=3, MOrderID=6\\
12-12,20:50 & ReceiveMO   & supplier=supp5, items=(it2), POrderID=3, MOrderID=6\\
12-13,03:20 & ReceiveSupplResp    & supplier=supp5, items=(it2), POrderID=3, MOrderID=6, answer=accept\\
12-13,03:21 & ReceiveItems       & supplier=supp5, items=(it2), POrderID=3, MOrderID=6\\
12-13,04:30 & ClosePO & POrderID=2\\
12-13,08:36 & Assemble & items=(it2), POrderID=3, MOrderID=6\\
12-13,08:37 & InvoicePO & POrderID=3\\
12-13,08:38 & ShipPO & POrderID=3\\
12-13,08:39 & ClosePO & POrderID=3
\end{tabular}}
    \caption{An example of a raw log.}
    \label{fig:RawLogExample}
\end{figure}
%
%
\section{Background}
\label{sec:background}
We first give the necessary background in order to present the artifact lifecycle discovery 
methods by a very brief introduction to the relevant modeling approaches.

\subsection{Petri nets}
\label{sec:PN}

Petri nets~\cite{petri-nets} are an established notation for modeling and analyzing workflow processes. 
Its formal basis allow to perform formal analysis w.r.t. many static and dynamic properties.
Petri nets are expressive while still being executable which makes them appropriate for application
in realistic scenarios.
They have been used in a wide variety of contexts and a great number of the developed process mining
techniques assume or generate Petri nets. 

A Petri net is a directed bipartite graph with two types of nodes called \emph{places} (represented by circles)
and \emph{transitions} (represented by rectangles) connected with arcs. Intuitively, the transitions
correspond to activities while the places are conditions necessary for the
activity to be executed. Transitions which correspond to business-relevant activities observable in the
actual execution of the process will be called visible transitions, otherwise they are
invisible transitions. A labeled Petri net is a net with a labeling function that assigns a label (name) for each place and transition. 
Invisible transitions are labeled by a special label $\tau$. 

An arc can only connect a place to a transition or 
a transition to a place. A place $p$ is called a pre-place of a transition $t$ iff there
exists a directed arc from $p$ to $t$. A place $p$ is called a post-place of transition
$t$ iff there exists a directed arc from $t$ to $p$. Similarly we define a pre-transition and 
a post-transition to a place.

At any time a place contains zero or more tokens, represented graphically as black dots. The current state of
the Petri net is the distribution of tokens over the places of the net. 
A transition $t$ is enabled iff each pre-place $p$ of $t$ contains at least
one token. An enabled transition may fire. If transition $t$ fires, then $t$ consumes one token
from each pre-place $p$ of $t$ and produces one token in each post-place $p$ of
$t$.

A Petri net $N$  is a \emph{workflow net} if
it has a distinguished initial place, that is the only place with no incoming
arcs, a distinguished final place, that is the only place with no outgoing
arcs, and if every transition of the net is on a path from initial to final
place. The initial place is also the only place with a token in the initial
marking. 

$N$ is \emph{free-choice} iff there is a place $p$ that is a
pre-place of two transitions $t$ and $s$ of $N$ ($t$ and $s$ compete for
tokens on $p$), then $p$ is the only pre-place of $t$ and of $s$. In a
free-choice net, a conflict between two enabled transitions $t$ and $s$ can
be resolved locally on a single place $p$. $N$ is \emph{sound} iff every run
of $N$ starting in the initial marking can always be extended to a run that
ends in the final marking (where only the final place of $N$ is marked), and
if for each transition $t$ of $N$ there is a run where $t$ occurs.

In order to model interactions between artifacts in artifact-centric systems, we can use 
\textbf{Proclets}~\cite{AalstBEW:2001:proclets} notation where each proclet 
represents one artifact type as a Petri net and constructs such as ports and channels
can be used to represent different types of interactions between the artifacts. In this paper
we concentrate on the artifact lifecycles. The difference to classical process discovery is that 
classical discovery considers just one monolithic 
process model (Petri net), whereas we consider sets of related Petri nets.  


Fig.~\ref{fig:ordertocash} shows one way of modeling the lifecycle of the Material order 
artifact type from the build-to-order example using Petri net notation.



\subsection{Guard-Stage-Milestone meta-model}
\label{sec:GSM}

The Guard-Stage-Milestone meta-model~\cite{gsm2, gsm1} provides a more declarative approach for modeling
artifact lifecycles which allows a natural way for representing hierarchy and parallelism within the same instance of
an artifact and between instances of different artifacts. 

The key GSM elements for representing the artifact lifecycle are stages, guards and milestones
which are defined as follows.

Milestones correspond to business-relevant operational objectives, and are achieved (and
possibly invalidated) based on triggering events and/or conditions
over the information models of active artifact instances.
Stages correspond to clusters of activities preformed for, with or by an artifact instance 
intended to achieve one of the milestones belonging to the stage. 

Guards control when
stages are activated, and, as with milestones, are based on triggering
events and/or conditions. A stage can have one or more guards and one or more milestones. 
It becomes active (or open) when a guard becomes true and inactive (or closed) when
a milestone becomes true.

Furthermore, sentries are used in guards and milestones, to control when stages
open and when milestones are achieved or invalidated.
Sentries represent the triggering event type and/or a condition of the guards and milestones.
The events may be external or internal,
and both the internal events and the conditions may refer to
the artifact instance under consideration, and to other artifact
instances in the artifact system.

And finally, stages can contain substages which can be open when the parent stage is active. 
If a stage does not contain any substages, it is called atomic.

\begin{figure}[t]\centering
\includegraphics[width=.60\textheight]{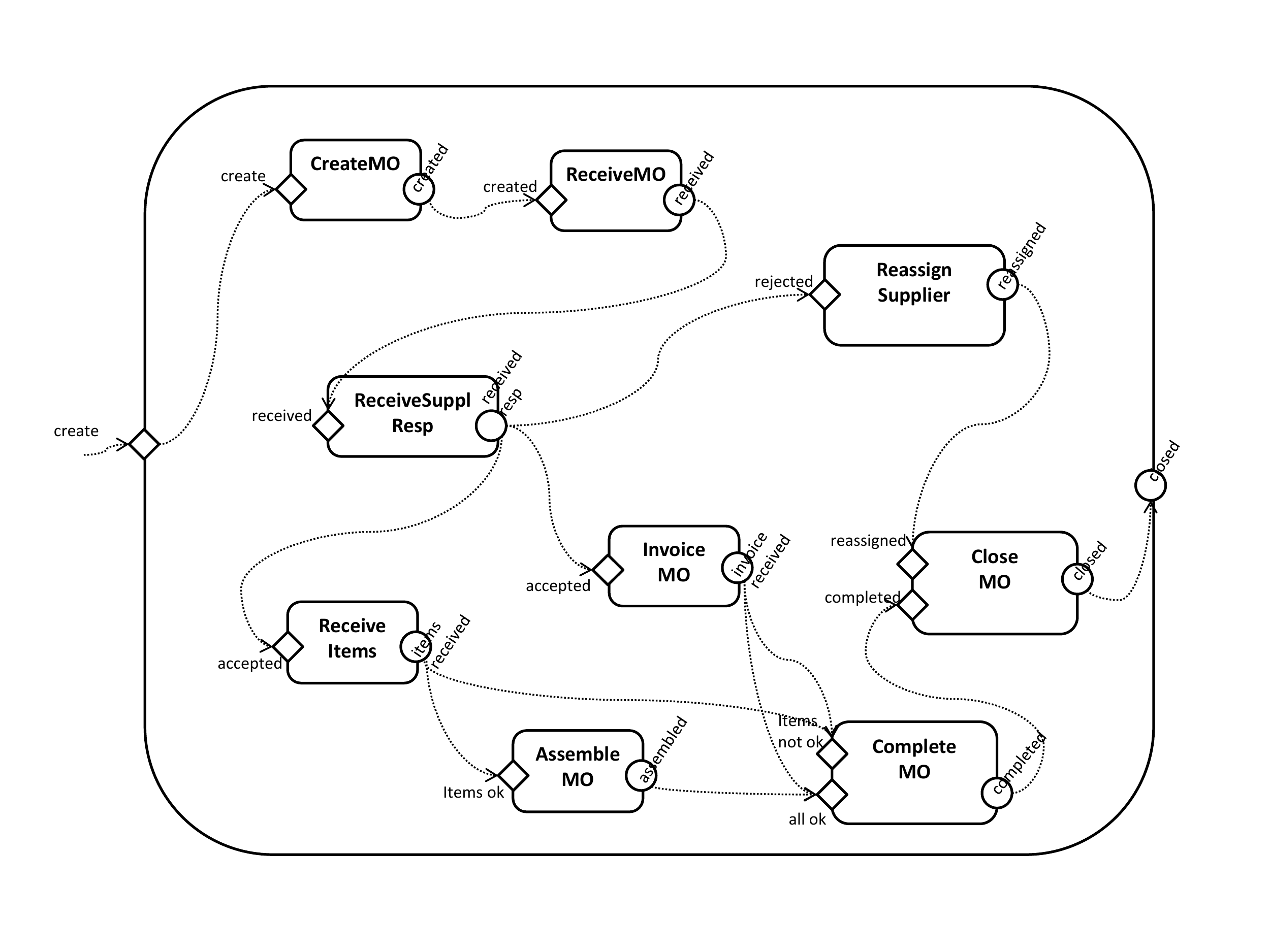}
\caption{A GSM model for the lifecycle of the Material Order artifact, part of the build-to-order example}
\label{fig:ordertocash-GSM}
\end{figure}

Fig.~\ref{fig:ordertocash-GSM} shows the Material order artifact type as a GSM model. The stages
are represented as rectangular shapes. A diamond on the left side of the stage represents a guard with
its name and a circle at the right side of the stage represents a milestone. The dotted lines
between stages are not part of the model and were only added to give an indication of the control
flow that is implicit through the guards and milestone sentries which are expressed in GSM in terms of logics.
%
%
%
%

%
\section{Artifact Structure Discovery}
\label{sec:art-struct}
The first problem to cope with when discovering models of artifact-centric processes 
is to identify which artifacts exist in the system. Only then a lifecycle model can be identified for each artifact.

In order to discover the artifact structure and subsequently generate the artifact-centric logs, 
we explore the implicit information contained
in the data belonging to the events.
As a first step we apply data mining and analysis
methods to the raw log data to discover correlation information between the events which
allows to build the underlying Entity-Relationship (ER) model. This includes methods for
discovering functional and inclusion dependencies which allow us to discover which
event types belong to the same entity and how entities are related to each other.

Using the discovered ER model, we perform analysis which then suggests to the user which entities should be
chosen as artifacts and provides support in the selection process. 

Finally, for the entities designated as artifacts, 
we can extract artifact instance-specific traces which can be used to discover the lifecycle of each artifact. 
The collection of such traces is called \emph{artifact-centric logs}. 

\begin{figure}[h]
\centering
\includegraphics[height=0.5\textwidth]{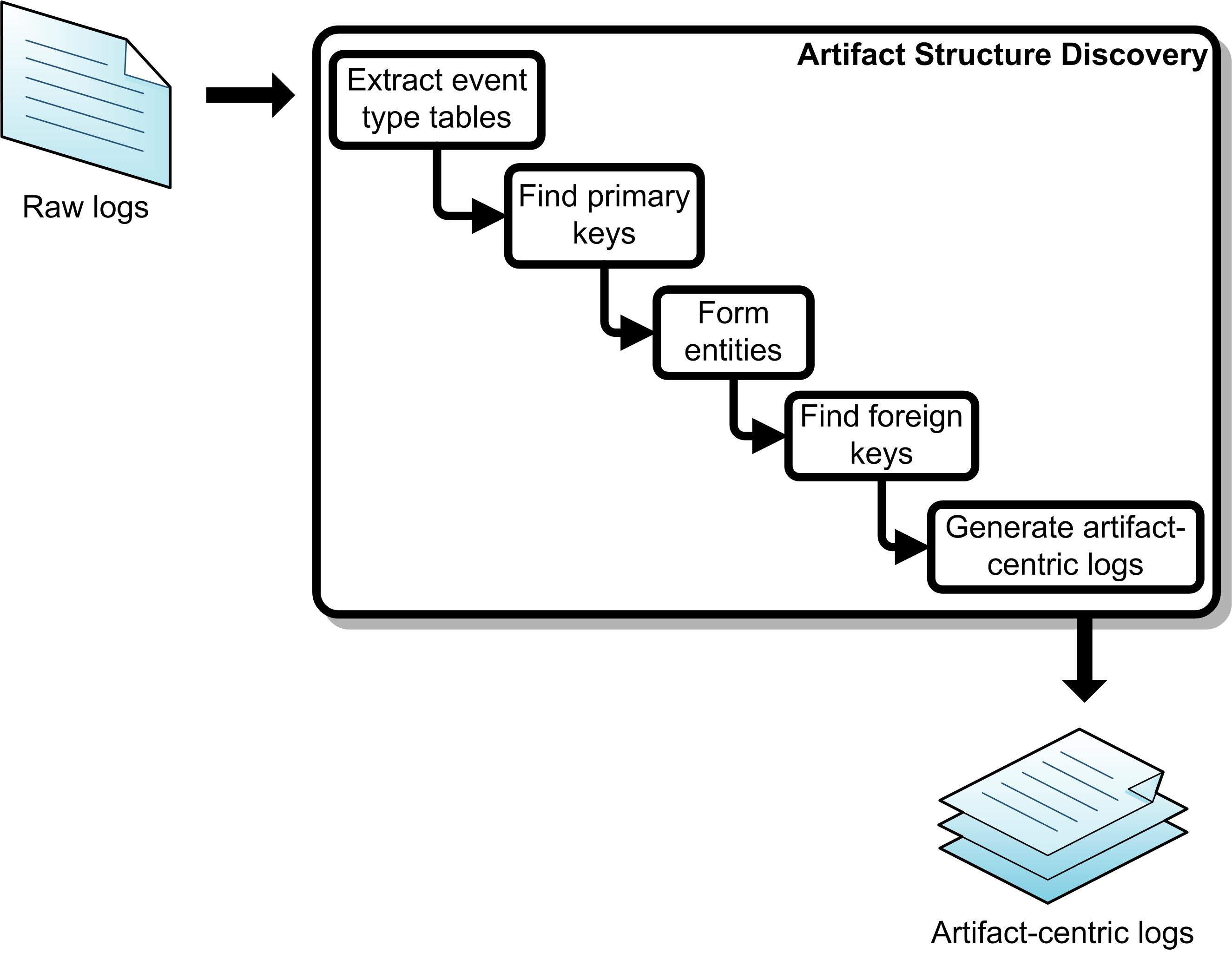}
\caption{Artifact structure discovery.}
\label{fig:art-struct}
\end{figure}

The rest of this section presents each of these steps in more detail, as also shown in Fig.~\ref{fig:art-struct}.

%
%
\subsection{Entity Discovery}
\label{sec:entity-discovery}
Each event in a raw log belongs to an event type, for example the event type $\fni{CreateMO}$ can have one or more instances in the log with specific
timestamps and for specific orders being created. Each event contains
a timestamp, one or more data attributes and belongs to exactly one event type. 
Formally, we define a raw log as described in the following paragraphs.

\begin{definition}[Event]\label{def:log:event}
Let $\{ A_1, A_2, \ldots, A_n \}$ be a set of attribute names and $\{ D_1, D_2, \ldots, D_n \}$ - a set of attribute domains
where  $D_i$ is the set of possible values of $A_i$ for $1\leq i\leq n$. 
Let $\Sigma = \{ a_1, a_2, \ldots, a_m \}$ be a set of event
types. Event $e$ is a tuple $e = (a, \tau, v_1, v_2, \ldots, v_k)$ where
\begin{enumerate}
\item $a \in \Sigma$ is the \emph{event type} to which $e$ belongs,
\item $\tau \in \Omega $ is the timestamp of the event where $\Omega $ is the set of all timestamps,
\item for all $1\leq i\leq k$ $v_i$ is an attribute-value pair $v_i=(A_i,d_i)$ where $A_i$ is an attribute name and $d_i\in D_i$
is an attribute value.
\end{enumerate}
All events of an event type $a$ are called event instances of $a$.
\end{definition}
Note that the definition allows for multiple attribute-value pairs of the same event to refer to the the same
attribute name. In such cases we talk about multi-valued attributes. For example event type $\fni{ReceivePO}$
has a multi-valued attribute $\fni{items}$.
%
%
\begin{definition}[Raw log]
A \emph{raw log} $L$ is a finite sequence of events $L = e_1 e_2 \ldots e_n$.
$L$ induces the total order $<$ on its events with $e_i < e_j$ iff $i < j$.
\end{definition}

The order of events in $L$ usually respects the temporal order of their timestamps, i.e. if event $e$ precedes event $e'$  temporally then $e < e'$ . 

To present the methods proposed in this Section, we adopt notation from
\emph{Relational Algebra}~\cite{SilberschatzKS:2001:database}. A \emph{table}
$T \subseteq D_1 \times \ldots \times D_m$ is a relation over domains $D_i$
and has a \emph{schema} $\Schema(T) = (A_1,\ldots,A_m)$ defining for each
\emph{column} $1 \leq i \leq m$ an \emph{attribute name} $A_i$. For each
\emph{entry} $t = (d_1,\ldots,d_m) \in T$ and each column $1 \leq i \leq m$,
let $t.A_i \df d_i$. We write $\Attr(T) \df \{A_1,\ldots,A_m\}$ for the
attributes of $T$, and for a set $\Tables$ of tables, $\Attr(\Tables) \df
\Union_{T \in \Tables}\Attr(T)$. The domain of each attribute can contain a
special value $\perp $ which indicates the null value, \ie the attribute is
allowed to have null values for some entries. The set of timestamps $\Omega $
does not contain the null value.

All instances of an event type form a data table where each row corresponds to an event
instance observed in the logs and consists of the values of all data attributes for that event. 
In order to transform this table into second normal form (2NF) we need to separate 
the multi-valued attributes and treat them differently - this will be discussed later.
In the following, a single-valued attribute $A$ for event type $a$ is an attribute
for which, for each event $e$ of type $a$ in raw log $\mathcal{L}$, $e$ contains
at most one attribute-value pair $(A_i,d_i)$ where $A_i=A$.

Formally, this can be expressed as follows.
\begin{definition}[Event type table]\label{def:event_type_table}
Let $a$ be an event type in the raw log $\mathcal{L}$ and let $E = \{ e_1, e_2, \ldots, e_n \}$ be the
set of events of type $a$, i.e. $e_i = (a, \tau_i, v_{i_1}, v_{i_2}, \ldots, v_{i_m})$ where $m\geq 1$ for all $1\leq i\leq n$
and $v_{i_j} = (A_j, d_{i_j})$ and $A_j$ is a single-valued attribute for $e_i$.
An \emph{event type table} for $a$ in $\mathcal{L}$ is a table $T\subseteq \Omega \times D_1\cup \{\perp \} \times \ldots \times D_m\cup \{\perp \}$ \st there exists an
entry $t = (\tau, d_1,\ldots,d_m) \in T$ iff there exists an event $e\in E$ where $e = (a, \tau, (A_1, d_1), (A_2, d_2), \ldots, (A_k, d_k))$
and $d_i\in D_i\cup \{\perp \}$ for $1\leq i\leq m$. The first attribute of the table is called the timestamp attribute
of the event type and the rest are the data attributes of the event type. 
\end{definition} 

Note that the events of the same type might in general have a different number of attributes and the schema of the event type table 
will consist of the union of all single-valued attribute names that appear in events of this type in the raw log. Therefore
there might be null values for some attributes of some events. 

Fig.~\ref{fig:event-type-tables} shows two of the event type tables for the raw log 
in Fig.~\ref{fig:RawLogExample}, namely for event types $\fni{ReceivePO}$ and $\fni{CreateMO}$. 

The set of event type tables for a given raw log forms a database which implicitly represents
information about a number of \emph{entities}. Such entities can be detected by discovering their identifiers in
the data tables of the event types. For example if the attribute $\fni{POrderID}$ is the identifier 
of both event types $\fni{ReceivePO}$ and $\fni{ShipProduct}$ then we can
assume that these two events refer to the same entity, in this case the entity $\fni{PurchaseOrder}$. 

These identifiers will be \emph{keys}
in the data tables and can be detected using algorithms for discovering functional dependencies between data attributes.
By selecting a primary key, the table is transformed into 2NF.

\begin{definition}[Key]\label{def:key}
A set of data attributes $\mathcal{A}=\{A_1, A_2, \ldots, A_n\}$ is a \emph{key}
in an event type table $T$ iff: 
\begin{enumerate}
\item for every data attribute $A'$ in $T$ and every pair of entries $t, t'\in T$ 
if $t.A'\neq t'.A'$ then $(t.A_1, t.A_2, \ldots, t.A_n)\neq (t'.A_1, t.A_2, \ldots, t.A_n)$, and
\item no subset of $\mathcal{A}$ is also a key.
\end{enumerate}
\end{definition} 

Note that the timestamp attribute of a table cannot be part of a key. 

For example $\{\fni{POrderID}\}$ is a key in the $\fni{ReceivePO}$ table in Fig.~\ref{fig:event-type-tables}
and $\{\fni{MOrderID}\}$
is a key in the table $\fni{CreateMO}$.

A table can have multiple keys from which only one is selected as the primary key. 
The multi-valued attributes can then be represented as additional data tables in the following way.
For each multi-valued attribute $A$ in event type $a$ with primary key $\{A_1, \ldots, A_n\}$
a new table $T$ is constructed, $T\subseteq D_1 \times \ldots \times D_n \times D$ such that:
$D$ is the domain of $A$, $D_i$ is the domain of $A_i$ for $1\leq i\leq n$ and for each tuple
$t\in T$, $t=(t.A_1, \ldots, t.A_n, t.A)$ there exists event $e = (a,\tau, (A_1,t.A_1),\ldots,(A_n,t.A_n),(A,t.A),\ldots)$.

An example of such an additional table is the table in Fig.~\ref{fig:additional-table:ReceivePO} which represents the 
multi-valued attribute $\fni{items}$ of event type $\fni{ReceivePO}$.
An event type table together with all the additional tables associated with it is called 
an event type cluster. 

Event type tables share keys if 
their primary keys have the same attribute names. An entity is a set of event type tables that share primary keys. 
The shared primary key is called an identifier for the entity. Every value of the identifier defines an instance
of the entity.

\begin{definition}[Entity, identifier, instance]\label{def:entity}
Let $\mathcal{T}=\{T_1,T_2, \ldots, T_m\}$ be the set of tables of a set of event type clusters in the raw log $\mathcal{L}$. 
Let $\mathcal{A}=\{A_1, A_2, \ldots, A_n\}$ be the (shared) primary key for the tables in $\mathcal{T}$. 
An \emph{entity} $E$ for $\mathcal{T}$ is defined as $E=(\mathcal{T}, \mathcal{A})$ and
$\mathcal{A}$ is referred to as an \emph{instance identifier} for $E$. An \emph{instance} $s$ of $E$ is defined as
$s=((A_1,d_1), \ldots, (A_n,d_n)$, $(A_{n+1},d_{n+1}), \ldots, (A_l,d_l))$ 
where an attribute-value pair 
$(A_i,d_i)$ is in $s$, ${n+1} \leq i\leq l$, iff there exists an entry $t\in T$, $T\in \mathcal{T}$, 
\st $t.A_i=d_i$ and for all $1\leq k\leq n$ $t.A_k=d_k$.
\end{definition}

In this context, we are not discovering and analyzing all entities in the sense of ER models.
The specific entities that we discover in the execution logs have behavior, and therefore a non-trivial lifecycle,
\ie go through different states.
The changes of state are represented by events in the logs which in turn are reflected in timestamp attributes.

For the raw log in Fig.~\ref{fig:RawLogExample}, we find two entities - one with a shared primary key $\{\fni{POrderID}\}$
and the other one with $\{\fni{MOrderID}\}$.

The primary key becomes the instance identifier for the corresponding entity. 
For example an instance of the entity with instance identifier $\fni{POrderID}=1$ will be
$s=((\fni{ReceivePO}, \tau_1 )$, $(\fni{ShipPO}, \tau_2)$, $(\fni{InvoicePO}, \tau_3)$, 
$(\fni{ClosePO},\tau_4))$ 
where $\tau_1$=11-24,17:12, $\tau_2$=11-25,12:11, $\tau_3$=11-26, 09:30 and $\tau_4$ = 12-03,14:34. Names for the entities
can be assigned based on the names of the attributes in their instance identifiers or 
more meaningful names can be given by the user if needed.

\begin{figure}\centering
{\scriptsize\sffamily
\begin{tabular}{|c|c|} \hline
\textbf{POrderID}     & ReceivePO      \\ \hline
$1$       & 11-24,17:12 \\
$2$       & 11-25,08:53 \\
$3$       & 12-04,15:03 \\ \hline
\end{tabular}

\begin{tabular}{|c|c|c|c|} \hline
\textbf{MOrderID}     & POrderID & supplier & CreateMO      \\ \hline
$1$       & 1 & supp6 & 11-24,17:13 \\
$2$       & 2 & supp1 & 11-26,09:31 \\
$3$       & 2 & supp4 & 11-28,12:22 \\ 
$4$       & 3 & supp2 & 12-04,15:33 \\ 
$5$       & 3 & supp5 & 12-04,16:30 \\ 
$6$       & 3 & supp5 & 12-12,20:41 \\ \hline
\end{tabular}
} \caption{Event type tables for $\fni{ReceivePO}$ and $\fni{CreateMO}$ 
discovered from the raw log in Fig.~\ref{fig:RawLogExample}}
\label{fig:event-type-tables}
\end{figure}

\begin{figure}\centering
{\scriptsize\sffamily
\begin{tabular}{|c|c|c|} \hline
POrderID  	&	items       \\ \hline
$1$      &	$it0$ 	 \\
$2$      &	$it0$   \\
$2$      &	$it1$ 	 \\ 
$2$      &	$it2$	 \\
$2$      &	$it3$  \\ 
$3$      &	$it1$  \\ 
$3$      &	$it2$  \\ \hline
\end{tabular}

} \caption{The additional table for multi-valued attribute $\fni{items}$ of event type $\fni{ReceivePO}$
from the raw log in Fig.~\ref{fig:RawLogExample}}
\label{fig:additional-table:ReceivePO}
\end{figure}

The event type tables of the same entity can be combined into one table where attributes with the same name
become one attribute. The only exception is when data attributes of the same name which are not part of the primary key
have different values for the same value of the primary key. Then they should be represented by different attributes
so that both values are present in the joined table. This does not occur in the example of Fig.~\ref{fig:RawLogExample}.

An example of an entity table for entity $\fni{PurchaseOrder}$ is given in Fig.~\ref{fig:entity-tables:PurchaseOrder}.
In the following we refer to the combined entity table by the name of the entity.

\begin{figure}\centering
{\scriptsize\sffamily
\begin{tabular}{|c|c|c|c|c|} \hline
\textbf{POrderID}     & ReceivePO     & ShipPO & InvoicePO & ClosePO \\ \hline
$1$       & 11-24,17:12 & 11-25,12:11 & 11-26,09:30 & 12-03,14:34 \\
$2$       & 11-25,08:53  & 12-06,09:34 & 12-06,07:25 & 12-13,04:30 \\
$3$       & 12-04,15:02 & 12-13,08:38 & 12-13,08:37 & 12-13,08:39 \\ \hline
\end{tabular}
} \caption{The combined table for the entity $\fni{PurchaseOrder}$}
\label{fig:entity-tables:PurchaseOrder}
\end{figure}

Once the entities have been discovered then any relationships among them need to be identified. This can be done using algorithms
for detecting inclusion dependencies, which discover candidate foreign keys between data tables. From the set of candidate 
foreign keys the user selects the foreign keys which are meaningful in the specific application. 

A \emph{database} $\DB = (\Tables, K)$ is a
set $\Tables$ of tables with corresponding schemata $\Schema(T)$, $T\in\Tables$
\st their attributes are pairwise disjoint, and a \emph{key relation} 
$K \subseteq (\Attr(\Tables) \times \Attr(\Tables))^\natnum $.

The key relation $K$ expresses foreign-primary key relationships between the
tables $\Tables$: we say that $\big((A_1,A_1'),\ldots,(A_k,A_k')\big) \in K$
\emph{relates} $T \in \Tables$ to $T' \in \Tables$ iff the attributes
$A_1,\ldots,A_k \in \Attr(T)$ together are a \emph{foreign key} of $T$
pointing to the \emph{primary key} $A_1',\ldots,A_k' \in \Attr(T')$ of $T'$.

In our context we are interested in foreign-primary key relationships between entities. A foreign key in entity $E_1$ of entity $E_2$
is an attribute in $E_1$ which is a foreign key to the instance identifier in $E_2$.

A foreign key in entity $E_1$ which refers to entity $E_2$ indicates a
relationship between $E_1$ and $E_2$. The foreign keys provide us with the necessary correlation information
of how specific instances of one entity relate to specific instances of another entity.

The additional tables associated to event type tables contain foreign keys to the
primary key of the event type table - such foreign keys do not need to be discovered 
as they are already known and included by design. 

Closer examination of the foreign keys values allows us to find the multiplicities
of the discovered relationships.

The collection of entities and their relationships in a raw log $\mathcal{L}$
forms an ER model.

For instance, $(\fni{MaterialOrder.POrderID},\fni{PurchaseOrder.POrderID})$ is a primary-foreign key
relation from the table of entity $\fni{MaterialOrder}$ to the table of entity $\fni{PurchaseOrder}$.  

Applying the proposed method to the example raw log we discover an ER model similar to the one in Fig.~\ref{fig:ER-model} 
with the exception of a few of the attributes ($\fni{PurchaseOrder.CreateWO}$, $\fni{PurchaseOrder.DetermineSuppl}$, 
$\fni{MaterialOrder.InvoiceMO}$, $\fni{MaterialOrder.CompleteMO}$ and $\fni{MaterialOrder.CloseMO}$) 
since they were not present in the log. 

%
%
\subsection{From Entities to Artifacts}
\label{sec:entities-artifacts}
After discovering the underlying ER model, one additional step is needed in order to discover which entities will become artifacts.
Intuitively, an artifact can consist of one or more entities from which one is the main entity which determines which events belong to the same 
instance, i.e. the primary key of the main entity becomes the artifact identifier and all events with the same value of the artifact
identifier belong to the same instance of the artifact.

As with any modeling process, there is always a subjective element in deciding what is important and needs to be represented
more prominently in the model. Such decision is also domain-specific and depends on the goals and purpose of the model.
In the context of artifact-centric modeling, this applies to the decision which entities are important enough to be represented
as separate artifacts. However certain guidelines do exist and they were used here to provide heuristics for pruning the options that are not 
appropriate and for assisting in the selection process.

First of all, an artifact needs to have a lifecycle which means that at least one event type has to be associated with it. 

Secondly, m-to-n relations between entities signify that they should belong to different artifacts. For example 
if a purchase order can be realized by multiple material orders and a material order combines items from different purchase orders
then combining them in one artifact is not a good design solution - an instance of such an artifact will overlap with multiple other instances of the same artifact.

Finally, we use the intuition that an entity whose instances are created earlier than the related instances (through foreign key relations)
of another entity has a higher probability of being more important. If an entity is not preceded in this way by another one then
it should be chosen to become a separate artifact in the artifact-centric model. This for example represents the case where instances of 
one entity trigger the creation of (possibly multiple) instances of another entity. In the build-to-order example this is the case for
the PurchaseOrder since one purchase order triggers the creation of one or more material orders and material orders cannot exist if
no purchase order exists. 

These considerations will be defined more precisely in the rest of this Section.

Let $\mathcal{E}$ be the set of discovered entities. 
%
Let $E,E'\in \mathcal{E}$ be entities and let the tuple $((A_1,A'_1), (A_2,A'_2), \ldots, (A_n, A'_n))\in K$ be a
primary-foreign key relationship which defines a relationship between $E$ and $E'$. 

Given such a relationship, we say that an instance $s$ of $E$ identifies a set of instances $\{s'_1,s'_2,\ldots, s'_m\}$ of $E'$ if
$s.A_i = s'_j.A'_i$ for all $1\leq i\leq n$, $1\leq j\leq m$. Here $s.A$ denotes the value of attribute $A$
for the instance $s$.
If, for all instances $s$ in $E$ and for all primary-foreign key relations between $E$ and $E'$, 
$s$ identifies at most one instance of $E'$ then we say that $E$ uniquely
identifies $E'$.

The function $H: \mathcal{E}\rightarrow \mathcal{P}(\mathcal{E})$ denotes the logical horizon of an entity \cite{Feldman-86} 
where an entity $E_1$ is in the logical horizon of
another entity $E_2$ if $E_2$ uniquely (and transitively) identifies $E_1$. 
For example the entity  $\fni{MaterialOrder}$ uniquely identifies the entity $\fni{PurchaseOrder}$
while $\fni{PurchaseOrder}$ does not uniquely identify $\fni{MaterialOrder}$.
Therefore $\fni{PurchaseOrder}\in H(\fni{MaterialOrder})$ and $\fni{MaterialOrder}\notin H(\fni{PurchaseOrder})$.

For each instance of entity $E \in \mathcal{E}$, one attribute signifies the timestamp of the creation  of this instance - the earliest
event that refers to this instance, i.e. the earliest timestamp in the instance. Based on this, for each two instances of the same or different entities
we can find out which was created before the other one.

\begin{definition}[precedence, top-level entity]\label{def:struct:top}
An entity $E_1 \in \mathcal{E}$ \emph{precedes} an entity $E_2 \in \mathcal{E}$, denoted by $E_1\prec  E_2$, iff 
$E_2 \in H(E_1)$ and for each instance of $E_1$, its creation is always earlier than the creation of the corresponding instance of $E_2$.
An entity is a \emph{top-level entity} if it is not preceded by another entity.
\end{definition}

In the build-to-order example, $\fni{PurchaseOrder}$ is a top-level entity if the raw logs are sufficiently close to the description
of the process. Therefore it will become a separate artifact. $\fni{MaterialOrder}$ can also become an artifact if the user
considers it sufficiently important. For example an entity which has a lot of event types (i.e. a more elaborate lifecycle) might
be a better choice for an artifact than a smaller entity - still domain-specific aspects and the context of the modeling process
need to be considered as well. In this particular example it seems a good choice to make $\fni{MaterialOrder}$ an artifact as well.

After artifacts are selected, the remaining entities can be joined with artifacts with which they have a key-foreign key relation.
If multiple such artifacts exist, then input from the user can be provided to select the most appropriate choice. The resulting grouping 
of the entities is called an artifact view which is defined more precisely as follows:

\begin{definition}[Artifact View]
Let $\mathcal{D} = (\mathcal{T},K)$ be a database of event type tables. Let
$\{E_1,\ldots,E_n\}$ be the entities over $\mathcal{D}$. A set $\mathit{Art}
\subseteq \{E_1,\ldots,E_n\}$ of entities is an \emph{artifact} iff:
\begin{enumerate}

\item $\mathit{Art}$ contains exactly one designated main entity 
$\mathit{main}(\mathit{Art})$ whose identifier is designated as the
artifact instance identifier, 
\item $\mathit{Art}$ contains at most one top-level entity
    $\mathit{top}(\mathit{Art})$ and, if it exists,  $\mathit{top}(\mathit{Art})$ = $\mathit{main}(\mathit{Art})$.
\end{enumerate}
A set $\{ \mathit{Art}_1,\ldots,\mathit{Art}_k \}$ of artifacts is an
\emph{artifact view} on $\mathcal{D}$ iff there is an artifact for each
entity in the set of 
top-level entities $\{E_1,\ldots,E_m\}$,  $m \leq k$.
\end{definition}

%
%
\subsection{Artifact-Centric Logs}
\label{sec:instance_aware_log}
In order to be able to discover an artifact's lifecycle, we need an explicit representation of the executions 
of its instances based on the events in the raw log. For this purpose,
here we introduce the notion of an \emph{artifact-centric log}. 
As a result of the raw log analysis steps discussed in the previous subsections,
the data is now represented in a database in 2NF. A set of entities are represented in the
database where each entity has a corresponding table whose primary key is the identifier
of the entity and the table contains all the event types belonging to the entity. 
A set of artifacts are also defined where each artifact consists of one or more entities.
Each artifact has as identifier the identifier of one of its entities while the rest are 
connected through foreign key relations to this entity. In this way, for each artifact we
can determine which event types belong to it (the event types of its entities) 
and which specific events belong to the same
instance of the artifact (determined by the same value of the artifact identifier).

Given is a database $\DB$,
its set of artifacts 
and their instance identifiers $\mathcal{I} = \{\mathcal{A}_1, \mathcal{A}_2, \ldots, \mathcal{A}_n\}$.
In the following, $\mathcal{I(\mathcal{T})}$ denotes
the set of instance identifiers present in a set of tables $\mathcal{T}$ 
(and their underlying ER model). Each instance identifier $\mathcal{A}=(A_1, A_2, \ldots, A_m)$
has a domain $D_1 \times \ldots \times D_m$. For simplicity we denote by $\mathcal{I}(\mathcal{A})=\{\id_1, \ldots, \id_k\}$ the set of
unique values of the instance identifier $\mathcal{A}$ present in the database, 
$\id_i\in D_1 \times \ldots \times D_m$. This corresponds
to the set of unique instances of the entity for which $\mathcal{A}$ is the instance identifier.

\begin{definition}[Instance-aware events]\label{def:log:instance_aware1}
Let $\Sigma = \{ a_1, a_2, \ldots, a_n \}$ be a finite set of \emph{event
types} and $\mathcal{I}$ the set of values of instance identifiers. An \emph{instance-aware event} $e$ is a tuple
$e = (a, \tau, \id)$ where:
\begin{enumerate}
\item $a \in \Sigma$ is the \emph{event type},
\item $\tau \in \Omega $ is the timestamp of the event,
\item $\id \in \mathcal{I}$ is the \emph{instance} for which $e$ occurred.
\end{enumerate}
Let $\mathcal{E}(\Sigma,\mathcal{I})$ denote the set of all instance-aware
events over $\Sigma$ and $\mathcal{I}$. 
\end{definition}
%
%
\begin{definition}[Artifact case, lifecycle log]
An artifact case $\rho = \langle e_1, \ldots, e_r \rangle \in
\mathcal{E}(\Sigma, \mathcal{I})$ is a finite sequence of instance-aware
events that
\begin{enumerate}
\item occur all in the same instance defined by a value of an
    instance identifier $\mathcal{A} \in \mathcal{I}$, i.e., for all
    $e_i, e_j \in \rho$ holds $id_i = id_j$, and
\item are ordered by their stamps, i.e., for all $e_i, e_j \in \rho$, $i
    < j$ implies $\tau_i \leq \tau_j$.
\end{enumerate}
An artifact lifecycle log for a single artifact is a finite set $L$ of
artifact cases. An artifact lifecycle log for an artifact system is a set
$\{L_1,\ldots,L_n\}$ of lifecycle logs of single artifacts.
\end{definition}




We can now address the problem of extracting from a given database $D$ 
(containing event information from a raw log) and for a given artifact view $V$, 
an artifact-centric log that contains a life-cycle log for each artifact in $V$.

Technically, we need to correlate each event in a table belonging to the artifact 
to the right instance of that artifact. Once this is done, all events of an artifact 
are grouped into cases (by the values of the artifact's indentifier), 
and ordered by time stamp. As the instance identifier for an event can be stored 
in another table the event's time stamp, we need to join the tables of the artifact 
to establish the right correlation. This requires some notion from relational algebra that we recall first.

Relational algebra defines several
operators~\cite{SilberschatzKS:2001:database} on tables. In the following, we
use \emph{projection}, \emph{selection}, and the canonical crossproduct. For
a table $T$ and attributes $\{A_1,\ldots,A_k\} \subseteq \Attr(T)$, the
projection $\Proj_{A_1,\ldots,A_k}T$ restricts each entry $t \in T$ to the
columns of the given attributes $A_1,\ldots,A_k$. Selection is a unary
operation $\Sel_\varphi(T)$ where $\varphi$ is a boolean formula over atomic
propositions $A = c$ and $A = A'$ where $A,A' \in \Attr(T)$ and $c$ a
constant; the result contains entry $t \in \Sel_\varphi(T)$ iff $t \in T$ and
$t$ satisfies $\varphi$ (as usual). We assume that each operation
correspondingly produces the schema $\Schema(T')$ of the resulting table
$T'$.

We define the partial function $\mathit{Path} : 2^{\mathcal{A}(\mathcal{T})}
\times 2^{\mathcal{A}(\mathcal{T})} \not\to K^*$ that returns for two sets of
attributes the sequence of key relations needed to connect the attributes.
Technically, $\mathit{Path}( \mathcal{A}_s, \mathcal{A}_e ) =
(\mathcal{A}_1,\mathcal{A}_1')\ldots(\mathcal{A}_k,\mathcal{A}_k')$ where:
\begin{enumerate}
\item for all $i=1,\ldots,k$, $(\mathcal{A}_i,\mathcal{A}_i') \in K$ or
    $(\mathcal{A}_i',\mathcal{A}_i) \in K$, and
\item there exist tables $T_1,\ldots,T_k \in \mathcal{T}$ such that
    $\mathcal{A}_s,\mathcal{A}_1 \in \mathcal{A}(T_1),
    \mathcal{A}_k',\mathcal{A}_e \in \mathcal{A}(T_k)$ and for all
    $i=1,\ldots,k-1$, $\mathcal{A}_i' \in \mathcal{A}(T_i)$ and
    $\mathcal{A}_{i+1} \in \mathcal{A}(T_i)$.
\end{enumerate}

Let $T(\Path(A_s,A_e)) = \{ T_1, ... , T_k \}$ denote the tables of this path.

To relate the values of $A_s$ and $A_e$ to each other, we need to join the tables 
of the path that connects both attribute sets. Let $\Path(A_s, A_e)$ be the path 
from $A_s$ to $A_e$ and let $\{T_1, ... , T_k \} = T(Path(A_s, A_e))$ be the involved tables. 
Then $\mathit{Join}(\Path(A_s, A_e)) = \Sel_{\varphi} ( T_1 \times ... \times T_k )$ where 
$\varphi = \logAND_{(A,A') \in \Path(A_s, A_e)} (A = A')$ selects from the cross product 
$T_1\times ... \times T_k$ only those entries which coincide on all key relations.

$\Path$ defines for each event type associated with a specific artifact the 
path to the instance identifier attribute associated to this artifact. 
The definition does not impose any restrictions on these paths however in
practice it often makes sense to choose the shortest path between the tables.
Classical graph theory algorithms such as Dijkstra algorithm can be used for 
finding all shortest paths. 


For the sake of illustrating these definitions, let us assume that MaterialOrder was not
chosen as a separate artifact and was joined with PurchaseOrder through the foreign key
relation provided by the attributes $\fni{MaterialOrder.POrderID}$ and $\fni{PurchaseOrder.POrderID}$. 
Table~\ref{tab:log-extraction policy} presents the artifact view for the
artifact $\fni{PurchaseOrder}$ for this variation of our running example. 
\begin{table}[t]
  \centering
  \scriptsize
  $\Sigma_{\fni{PurchaseOrder}} = \{ \fni{ReceivePO}, \fni{CreateMO}, \fni{ReceiveMO}, \fni{ReceiveSupplResponse}, 
\fni{ReceiveItems}, \fni{Assemble}, \fni{ReassignSupplier},$ $\fni{ShipPO},\fni{InvoicePO}, \fni{ClosePO} \}$,\\
  $\Inst(\Sigma_{\fni{PurchaseOrder}}) = \fni{POrderID}$\\
  \begin{tabular}{|l|l|l|}
    \hline
    \textbf{event type} $a \in \Sigma_{\fni{PurchaseOrder}}$ & $\TS(a)$ & $\Path_i(a)$\\
    \hline
    \fni{ReceivePO}             & \fni{ReceivePO}      & \{(\fni{PurchaseOrder.POrderID},\fni{PurchaseOrder.POrderID})\} \\
    \fni{CreateMO}             & \fni{CreateMO}      & \{(\fni{MaterialOrder.POrderID},\fni{PurchaseOrder.POrderID})\} \\
    \fni{ReceiveMO}  & \fni{ReceiveMO}       & \{(\fni{MaterialOrder.POrderID},\fni{PurchaseOrder.POrderID})\}\\
    \fni{ReceiveSupplResponse}     & \fni{ReceiveSupplResponse}        & \{(\fni{MaterialOrder.POrderID},\fni{PurchaseOrder.POrderID})\}\\
    \fni{ReceiveItems} & \fni{ReceiveItems}      & \{(\fni{MaterialOrder.POrderID},\fni{PurchaseOrder.POrderID})\}\\
    \fni{Assemble}        & \fni{Assemble} & \{(\fni{MaterialOrder.POrderID},\fni{PurchaseOrder.POrderID})\}\\
    \fni{ShipPO}             & \fni{ShipPO}      & \{(\fni{PurchaseOrder.POrderID},\fni{PurchaseOrder.POrderID})\} \\
    \fni{InvoicePO}             & \fni{InvoicePO}      & \{(\fni{PurchaseOrder.POrderID},\fni{PurchaseOrder.POrderID})\} \\
    \fni{ClosePO}             & \fni{ClosePO}      & \{(\fni{PurchaseOrder.POrderID},\fni{PurchaseOrder.POrderID})\} \\
    \fni{ReassignSupplier}             & \fni{ReassignSupplier}      & \{(\fni{MaterialOrder.POrderID},\fni{PurchaseOrder.POrderID})\} \\
    \hline
  \end{tabular}
  \caption{The artifact view for artifact $\fni{PurchaseOrder}$}
  \label{tab:log-extraction policy}
\end{table}

%
\vspace{.5em}\noindent{}\textbf{Log Extraction.}
After specifying an artifact view, an artifact log can be extracted
fully automatically from a given database $\DB$.
%
%
%

Let $TS_i$ be the set of all timestamp attributes in $\mathcal{T}_i$, 
i.e. $TS \in TS_i$ iff there exists table $T \in \mathcal{T}_i$ and 
$TS \in \mathcal{A}(T)$ with domain of $T$ being $\Omega$.

\begin{definition}[Log Extraction]
Let $\mathcal{D} = (\mathcal{T},K)$ be a database and let
$\{E_1,\ldots,E_n\}$ be the entities over $\mathcal{D}$. Let $\{
\mathit{Art}_1,\ldots,\mathit{Art}_k \}$ be an \emph{artifact view} on
$\mathcal{D}$.

For each artifact $\mathit{Art}_i$, the artifact life-cycle log of
$\mathit{Art}_i$ is extracted as follows.
\begin{enumerate}
\item Let $(\hat{\mathcal{T}}_i,\hat{\mathcal{A}}_i) =
    \mathit{main}(\mathit{Art}_i)$ be the main entity of
    $\mathit{Art}_i$ with identifier $\hat{\mathcal{A}}_i$. Let
    $\mathcal{T}_i = \bigcup_{(\mathcal{T},\mathcal{A}) \in
    \mathit{Art}_i} \mathcal{T}$ be the set of all tables of
    $\mathit{Art}_i$.
\item For each timestamp attribute $TS \in TS_i$ we define the table $T_{TS}^+ =
    \mathit{Proj}_{\hat{\mathcal{A}}_i,\mathit{TS}}\mathit{Join}(\mathit{Path}(
    \{\mathit{TS}\}, \hat{\mathcal{A}}_i )$.
\item Each entry $t = (id, ts) \in T_{TS}^+$ defines an instance aware event
    $e = (TS,ts,id)$ of type with timestamp $TS$ in instance $id$.
\item The set of all instance-aware events of artifact $i$ is
    $\mathcal{E}_i = \{ (TS,ts,id) | \exists TS \in TS_i, (ts,id) \in T_{TS}^+ \}$, let
    $\mathcal{I}_i = \{ id \mid (TS,ts,id) \in \mathcal{E}_i\}$ be the
    instance identifiers of $\mathit{Art}_i$.
\item For each $id \in \mathcal{I}_i$, the artifact case $\rho_{id}$
    contains all events $(TS,ts,id) \in \mathcal{E}_i$ ordered by their
    timestamps.
\item The artifact log $L_i = \{ \rho_{id} \mid id \in \mathcal{I}_i\}$
    contains all artifact cases of $\mathit{Art}_i$.
\end{enumerate}
\end{definition}

%
\begin{table}\centering
{\scriptsize\sffamily
\begin{tabular}{|c|c|c|c|c|c|}
\multicolumn{6}{c}{$\mathit{Join}(\{\fni{PurchaseOrder}, \fni{MaterialOrder}\}, K)$} \\ \hline
PurchaseOrder    & MaterialOrder  & MaterialOrder       & MaterialOrder  & MaterialOrder    & \ldots    \\ 
.POrderID 		& .POrderID 		& .supplier		& .ReceiveMO 		& .MOrderID &  \\ \hline
$1$       & $1$         & supp6    & 11-24,19:56   & $1$ & \ldots  \\
$2$       & $2$        & supp1 & 11-28,08:12     & $2$  & \ldots \\
$2$       & $2$         & supp4 & 12-03,14:54      & $3$ & \ldots  \\
$3$       & $3$       & supp2   & 12-04,15:56       & $4$ & \ldots \\
$3$       & $3$         & supp5 & 12-05,09:32         & $5$     & \ldots \\
$3$       & $3$         & supp5 & 12-12,20:50       & $6$      & \ldots \\ \hline
\end{tabular}}
\caption{Intermediate table when extracting events of artifact
$\fni{order}$.} \label{fig:extraction:order}
\end{table}

We illustrate the log extraction using our running example from
Sect.~\ref{sec:build-to-order}. We consider the artifact view on
$\fni{PurchaseOrder}$ as specified in Tab.~\ref{tab:log-extraction policy}. We
explain event extraction on event type $\fni{ReceiveMO}$.
\begin{enumerate}
\item First join the tables $\fni{PurchaseOrder}$ and $\fni{MaterialOrder}$ on \\
    $(\fni{PurchaseOrder.POrderID},\fni{MaterialOrder.MOrderID})$; Tab.~\ref{fig:extraction:order}
    shows a part of the resulting table.
\item Projection onto the instance identifier $\Inst(\fni{PurchaseOrder}) = \fni{POrderID}$ and
    the timestamp attribute $\fni{ReceiveMO}$ yields
    six entries $(1,\fni{11-24,19:56})$, $(2,\fni{11-28,08:12})$, $(2,\fni{12-03,14:54})$ and so on. 

\end{enumerate}
Similarly, events for the other event types of artifact $\fni{PurchaseOrder}$ can be
extracted to construct full cases.

In general we extract one log for every artifact and they can be considered independently to
discover the lifecycle of each artifact. This will be discussed in the next Section.
%
%
%
\section{Artifact Lifecycle Discovery}
\label{sec:lifecycle-disc}
Using the artifact-centric logs, we apply process
mining techniques in order to discover the lifecycle of each artifact independently. 
A great number of algorithms for process discovery exist with varying representational and search bias~\cite{pm-book,pd-survey}. 
The representational bias refers to the chosen formalism for representing the process model. Most approaches
use some form of a directed graph, for example Petri Nets~\cite{ga2,hierarchy1,alpha,heuristic}, Finite State Machines~\cite{CookWolf95}, 
Causal Nets~\cite{causal-nets}, Process Trees~\cite{tree-miner}, 
and so on. A few use other representations such as temporal logic~\cite{declarative1} or user-defined constraint templates~\cite{declarative2}.

The search bias refers to the algorithm used to traverse the solution space and the criteria used to select the final answer
(i.e., the generated process model). Many approaches have been proposed including Markov Models~\cite{CookWolf95}, 
Genetic Algorithms~\cite{ga1,ga2,tree-miner}, Neural Networks~\cite{CookWolf95}, Integer Linear Programming~\cite{ilp}, 
custom algorithms~\cite{alpha,hierarchy1,heuristic,causal-nets} and so on. 

The primary goal of the process discovery approaches is to generate models that accurately represent the behavior of
the system as evidenced by the logs. A number of criteria were defined as well as measures for assessing to what degree
the model reflects the desired behavior, as discussed in~\cite{pm-book}. 
The four prominent measures are fitness (to which degree a model can replay each trace in the log), 
generalization (to which degree a model allows replaying traces that are similar to the traces in the log 
and considered as part of the process), precision (to which degree a model allows replaying traces 
that are not contained in the log, but different from the traces in the log and not considered part of the process), 
and structural simplicity of the model. These four measures are partly contradictory, 
e.g. the simplest model has a very low precision, a perfectly fitting model can require very complex structure 
and thus a low simplicity. For this reason, different algorithms consider different subsets of these criteria. 

For complex processes, the importance of the readability (i.e. simplicity) measure for the process model increases. Different
approaches to increasing the readability of the generated models have been proposed including model simplification~\cite{simplification}, abstraction~\cite{hierarchy1}, 
fuzzy models~\cite{fuzzy1,fuzzy2}, trace clustering to generate simpler process variants~\cite{clustering},
block-structured models~\cite{tree-miner} and so on.

As an illustration, we briefly describe two of the successful algorithms with their advantages and disadvantages.

\begin{figure}[h]
\centering
\includegraphics[height=0.7\textwidth]{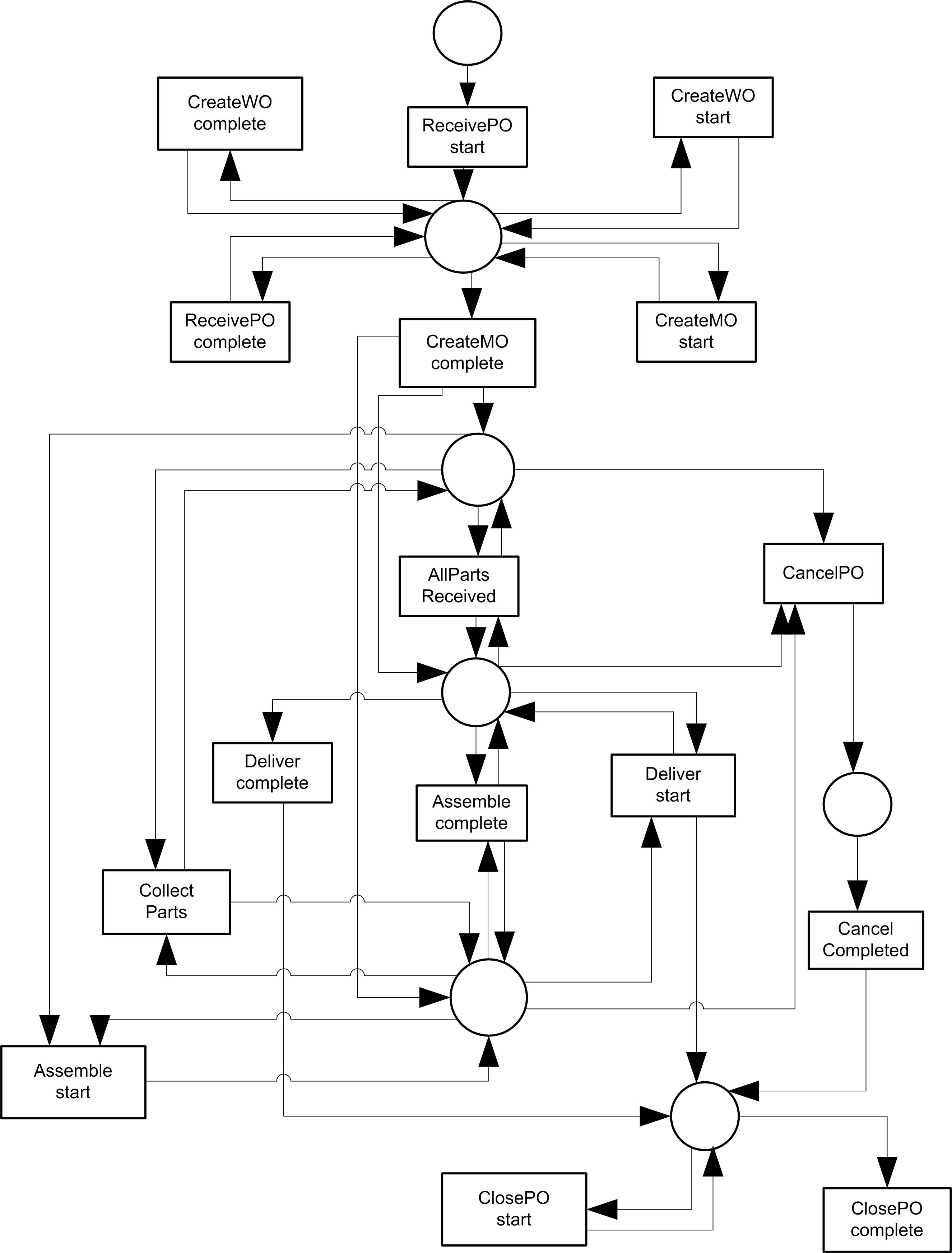} 
\caption{Petri Net mined using the ILP Miner.}
\label{fig:ilp}
\end{figure}

The ILP Miner~\cite{ilp} uses an Integer Linear Programming approach based on the language-based theory of regions. 
It generates models that are guaranteed to be consistent with the logs. However it can produce models that are not
very structured and less readable (i.e. spaghetti models). As an example, Figure~\ref{fig:ilp} shows a Petri net model mined
from logs for the $\fni{PurchaseOrder}$ artifact in a variation of the build-to-order example. 

The Tree Miner~\cite{tree-miner} uses a Genetic Algorithm for generating a model where the fitness function
is balanced between the four criteria for model quality: replay fitness, precision, simplicity and generalization. The
internal representation of the model is in the form of a process tree which has activities as leaves and the internal nodes
represent operators such as sequence, parallel execution, exclusive choice, non-exclusive choice and loop execution.
This guarantees that the generated models will be block-structured and sound. However the models are not guaranteed
to have perfect fitness and therefore might not be consistent with the logs. Another disadvantage is that the
algorithm takes significantly more time than for example the ILP miner. 

\begin{figure}[h]
\centering
\includegraphics[height=0.99\textwidth]{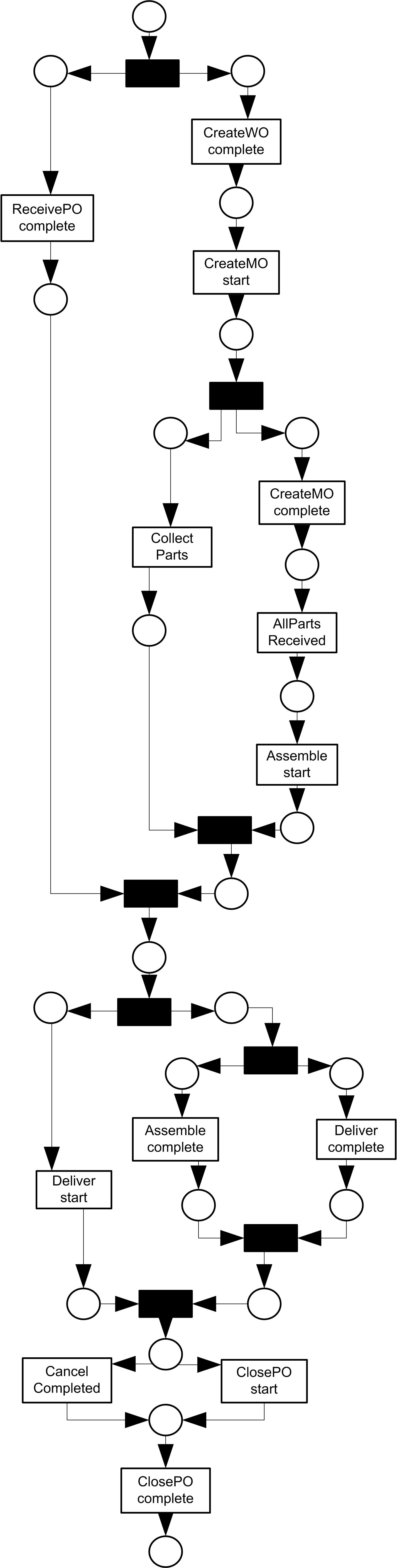} 
\caption{Petri Net mined using the Tree Miner.}
\label{fig:tm}
\end{figure}

Figure~\ref{fig:tm} shows a Petri net model mined from
the same logs of the $\fni{PurchaseOrder}$ artifact using the tree miner. 
We can clearly see that the model is more structured and readable than the one in
Figure~\ref{fig:ilp}. The downside is that this model deviates somewhat from the logs - for example a couple of the activities that appear in the logs
are not present in the model. Running the algorithm multiple times until a better fitness is achieved could potentially 
result in a better model though this will additionally increase the necessary time. 

In addition to the wide range of process discovery algorithms, a number of other relevant approaches have been proposed
for tasks such as conformance checking~\cite{conformance2,conformance1}, model repair~\cite{repair} and so on. 
Conformance checking methods can be used in this setting
to check if the generated model conforms to the logs and, if that is not the case, model repair methods can be applied
to repair the misconformances.

The above overview proves the benefits of choosing an approach that allows to reuse existing work and allow flexibility
and compositionality of the tool chain. Since the majority of process discovery and analysis approaches generate Petri nets
or models that can trivially be converted to Petri nets, we choose Petri nets as intermediate representation of the single artifact
models generated from artifact-centric logs.
The generated Petri nets are then translated into Guard-Stage-Milestone
models which represent the same behavior. Next Section presents the proposed translation algorithm in detail.

%
%
\section[Petri Nets to GSM models]{Petri Nets to GSM models}
\label{sec:PN2GSMgen}
We now concentrate on the last step in our artifact lifecycle discovery tool chain which deals with
translating Petri net models to GSM. 

As discussed in Section~\ref{sec:background}, in the GSM model each (atomic) stage has a guard and a milestone. 
The stage becomes active if and when the guard sentry evaluates to true. 
The guard sentry depends on a condition and/or an event. Events can be internal or external and can reflect
the changes in the state of the instance or other instances, for example opening or closing of stages, achieving
of milestones, etc. When an atomic stage opens, the activity associated with it will be executed. This can happen
automatically (for example when the activity is the assignment of a new value(s) to variable(s) in the instance's
information model) or can require actions by human agents. In the latter case the actual task execution is
external. The finishing of the execution of the task also generates an event which is denoted here af follows:
for an atomic stage $A$, the event $\fni{ATaskExecuted()}$ is generated when the task execution finishes. 
The milestone of the stage is achieved or invalidated also based on a sentry that depends on an event and/or condition. 
Achieveing a milestone $M$ of some stage generates an event $\fni{MAchieved()}$ that can be used in the guard of another stage. 
This way, the execution of stages can be ordered.

A straightforward approach to translating Petri nets to GSM models would proceed as follows. 
The visible transitions of the Petri net represent activities which are part of the business process. 
Therefore it is logical to represent them as atomic stages where the activity corresponds 
to the task associated with the stage. The control flow of the Petri net can then be encoded using the 
guards and milestones of these stages. 

More specifically, in the Petri net $N$, the order of execution of transitions is expressed by
the places and arcs of $N$. To impose the same execution order on the stages
of $M$, we encode the execution order of $N$ in the guards and milestones of
the stages of $M$. In particular, if transition $t_2$ is a successor of
transition $t_1$, then the guard $g_{t_2}$ has to express the opening of
stage $s_{t_2}$ in terms of the milestone $m_{t_1}$ of stage $s_{t_1}$. 

It is possible to encode the places of the Petri net (and their marking) in variables
which will be part of the information model of the artifact. These 
variables will be assigned true or false simulating the presence or absence of tokens in the places. 
This will be a relatively intuitive approach for designers skilled in the Petri net notation. 
However we argue that this would make the model less intuitive to the user and the 
relations between the tasks and stages become implicit and not easy to trace. Here we take a different 
approach which will be discussed first at a more general level and in the next subsections in more detail.

The intuition behind this approach is that the immediate ordering relations between transitions in the Petri net are extracted, 
translated into conditions and combined using appropriate logical 
operators (for AND- and XOR-splits and joins) into sentries which are then assigned to the guards. 
The milestones are assigned sentries that depend on the execution of the task associated 
with the stage - a milestone is 
achieved as soon as the task is executed and is invalidated when the stage is re-opened. 

As an example, consider the transition  $\fni{ReceiveMO}$ from the build-to-order model in Fig.~\ref{fig:ordertocash}. 
It can only be executed after the transition  $\fni{CreateMO}$ has been executed 
and there is a token in the connecting place. This can be represented as a part of a GSM model in the following way. Both transitions are represented by atomic stages. 
The guard of the stage  $\fni{ReceiveMO}$ has a sentry with expression  
$\fni{``on CreateMOMilestoneAchieved()"}$ where $\fni{CreateMOMilestoneAchieved}$ is the name
of the event generated by the system when the milestone of stage $\fni{CreateMO}$ is achieved. 
Therefore the sentry will become true when the event of achieving
the milestone of stage  $\fni{CreateMO}$ occurs and the stage will be open. 

The milestone of stage  $\fni{ReceiveMO}$ 
has a sentry $\fni{``on ReceiveMOTaskExecuted()"}$ where $\fni{ReceiveMOTaskExecuted}$
is the name of the event generated by the system when the task associated with atomic stage 
$\fni{ReceiveMO}$ completes. Therefore the sentry will become true when the associated task is executed,
 the milestone will be achieved and the stage - closed. 
Similarly the milestone of  $\fni{CreateMO}$ has a sentry $\fni{``on CreateMOTaskExecuted()"}$.

While this example is very straightforward, a number of factors can complicate the sentries. 
For example, we need to consider the possibility of revisiting a stage 
multiple times - this can be the case when the corresponding transition in the Petri net is part of a loop. 
At the same time, the transition might depend on the execution of multiple 
pre-transitions together and this cannot be represented using events - conditions need to be used instead. The conditions should express the fact that new executions of 
the pre-transitions 
have occurred. This means that the last execution of each relevant pre-transition 
occurred after the last execution of the transition in focus but also after every ``alternative" transition,
i.e., transition that is an alternative choice.

For example consider the transition  $\fni{CompleteMO}$ in Fig.~\ref{fig:ordertocash} which can fire 
for example if both  $\fni{AssembleMO}$ and  $\fni{InvoiceMO}$ have fired
(which will result in tokens in both pre-places). 
While this is not part of the model, imagine the hypothetical situation 
that  $\fni{CompleteMO}$,  $\fni{AssembleMO}$ and  $\fni{InvoiceMO}$ were part of a loop and could be executed multiple times. 
Since a sentry cannot contain multiple events, the guard of  $\fni{CompleteMO}$ has to be
expressed by conditions instead. The na\"{\i}ve solution ``if $\fni{AssembleMOTask.hasBeenExecuted}$ and $\fni{InvoiceMOTask.hasBeenExecuted}$" 
which checks if the two tasks have been executed in the past is not correct, since it becomes true the first time
the activities  $\fni{AssembleMO}$ and  $\fni{InvoiceMO}$ were executed and cannot reflect any new executions after that. 
We need a different expression to represent that new executions have occurred that have not yet 
 triggered an execution of  $\fni{CompleteMO}$. This will be discussed in detail in the next section.
 

Another factor that needs to be considered is the presence of invisible transitions, i.e., 
transitions without associated activity in the real world. For such invisible transitions no stage 
will be generated as they have no meaning at the business level that is meant to be reflected in the GSM model. 
Therefore, in order to compose the guard sentries, only visible pre-transitions 
should be considered. Thus we need to backtrack in the Petri net until we reach a visible 
transition and ``collect" the relevant conditions of the branches we traverse. 
As an example, consider the transition  $\fni{ReceiveItems}$ in Fig.~\ref{fig:ordertocash}. 
It can only fire when  
the invisible pre-transition represented by a black rectangle fires. We backtrack to find
the pre-places of the invisible transition and their pre-transitions. Here we determine that the only such pre-transition is 
 $\fni{ReceiveSupplResponse}$ and this branch has an associated condition - we can only take 
this branch if the supplier rejects an order and a new supplier has to be determined.  

With all these considerations in mind, the resulting guard sentry can become more complex and partly lose its advantage of being able to give intuition about 
how the execution of one task influences the execution of 
others. In order to simplify the sentry expressions, we apply methods for decomposing the expression 
into multiple shorter and more intuitive sentries which are then assigned to separate guards of the same stage.
Each guard of a stage forms an \emph{alternative} way of opening a stage, i.e., only one guard has to evaluate to true.
The composition and decomposition of guard sentries will be described more precisely in the next section. 

\begin{figure}[h]
\centering
\includegraphics[height=0.4\textwidth]{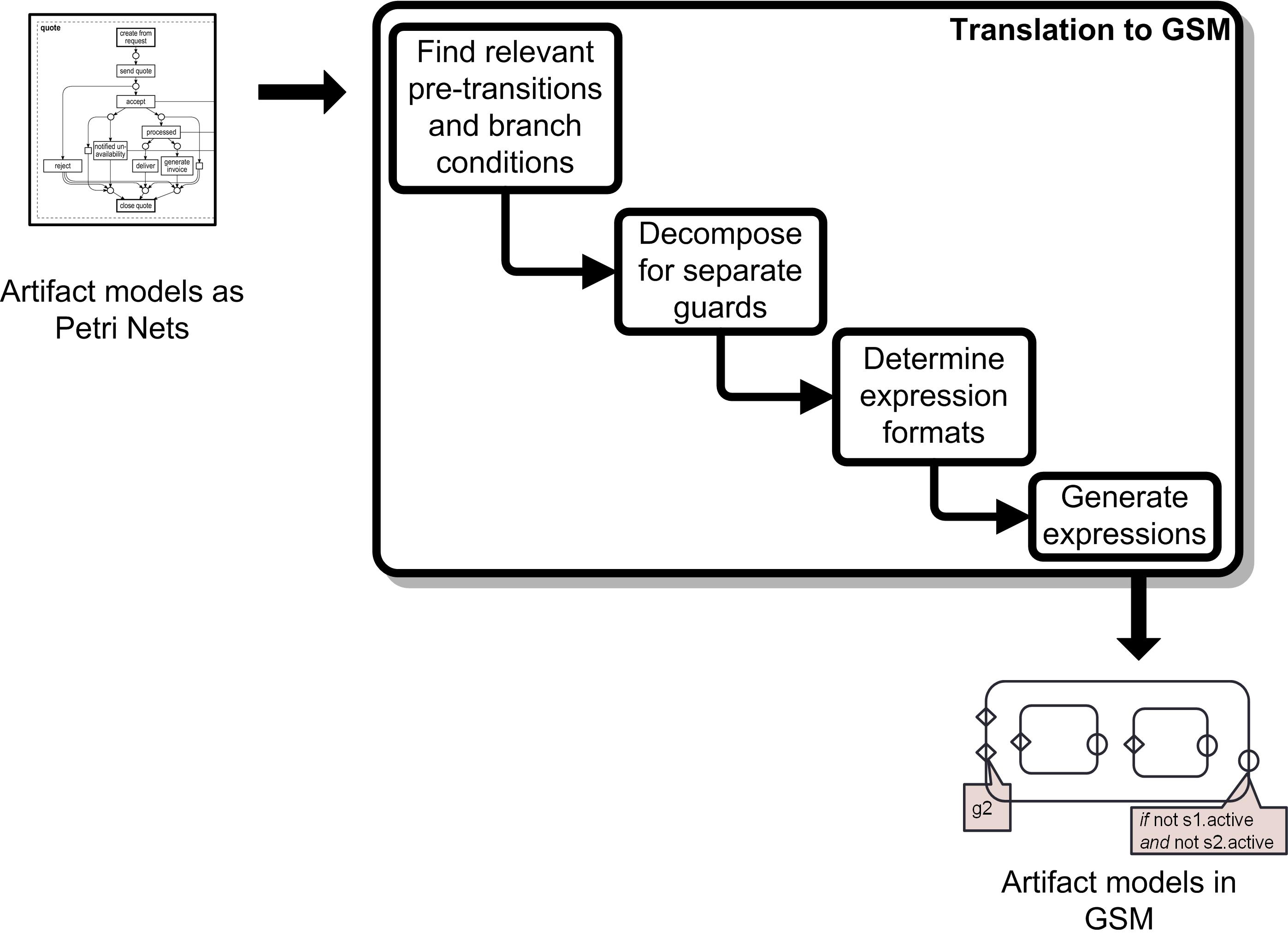}
\caption{Translation of Petri Nets to GSM models.}
\label{fig:translation}
\end{figure}

Let $t_o$ be the ``origin", i.e., the (visible) transition for which we compose a guard. 
At a more abstract level the proposed method for generating guard sentries for the stage of $t_o$ proceeds as follows
(as also shown in Fig.~\ref{fig:translation}):

Step 1: Find the relevant branch conditions and the pre-transitions whose execution will (help) trigger the execution of $t_o$.

Step 2: Decompose into groups that can be represented by separate guards.

Step 3: For each group, determine the appropriate format of the sentry and generate its expression.

We describe in detail each of these steps in the next subsections.

%
%
\subsection{Guard Sentries Generation}
\label{sec:PN2GSMguards}
Our approach for achieving step 1 is inspired by the research presented in~\cite{spm-to-bpel} for translating 
BPMN models and UML activity Diagrams into BPEL. 
It generates so-called precondition sets for all activities which encode possible ways of
enabling an activity. Next, all the precondition sets with their associated activities, are transformed into a set of Event-Condition-Action (ECA) rules.

Before giving the precise definitions of the approach proposed here, we first illustrate the intuition behind it by a couple of examples. 
Consider the transition  $\fni{CompleteMO}$ in Fig.~\ref{fig:ordertocash}. In order for it to be enabled and subsequently fire,
there need to be tokens in both of its pre-places. Therefore the precondition for enabling  $\fni{CompleteMO}$ is a conjunction of
two expressions, each of which related to one pre-place and representing the fact that there is a token in this pre-place.
This token could come from exactly one of the pre-transitions of this place. 

As a second example, consider the transition  $\fni{CloseMO}$.
It has one pre-place which has two pre-transitions. The token needed to enable $\fni{CloseMO}$
could come from either $\fni{CompleteMO}$ or $\fni{ReassignSupplier}$.
Therefore the precondition here is a disjunction of two expressions each related to
the firing of one pre-transition. 

Thus the general form of the composed guard sentry expression is a conjunction of disjunctions of expressions. These expressions, however, 
can themselves be conjunctions of disjunctions. This happens when a pre-transition is invisible (not observable in reality) and we need to consider recursively
its pre-places and pre-transitions. 

The building blocks of the composed expression are expressions each of which corresponds to the firing
of one visible transition $t$ that can (help) trigger the firing of the transition in focus $t_o$ (the ``origin"). 
We denote each of these building blocks by $\mathit{tokenAvailable}(t,t_o)$ for a transition $t$ with respect to $t_o$ 
and they will be discussed in the next section. They represent the intuition that $t$ has produced a token
that can enable $t_o$ if all other necessary tokens (produced by other pre-transitions) are available.
More precisely, $\mathit{tokenAvailable}(t,t_o)$ is true if $t$ has produced a token that has not yet 
been consumed by $t_o$ or by any other transition that could be enabled by it and is false otherwise.

The presence of a token in a pre-place is not a guarantee that a transition will fire. 
In the case of  $\fni{AssembleMO}$, a token in its pre-place enables two transitions,  $\fni{AssembleMO}$ and
 an invisible transition, but only one will fire. In order to resolve the non-determinism w.r.t. which transition
fires and consumes the token, conditions are associated with each
 outgoing arc of the place. Here $\fni{AssembleMO}$ will fire if the received items are of sufficient quality. 
These conditions are domain-specific and, in the following, we assume that these conditions are given - 
they can be provided by the user or mined from the logs using existing tools such as the decision miner from~\cite{dec-miner}. 
Therefore, the general form of the composed expression should have these conditions added to the conjunction.

Using this approach for the transition $\fni{CompleteMO}$ we can find a sentry expression of the following type
(only given informally here):
``$\fni{CompleteMO}$ will open if $\fni{InvoiceMO}$ is executed AND ($\fni{AssembleMO}$ was executed OR 
($\fni{ReceiveItems}$ was executed AND quality is insufficient))". In the rest of this Section the format 
will be defined precisely in a way that allows automatic generation given a Petri net model.

For the translation, we assume that the artifact lifecycle is modeled as a
sound, free-choice workflow net $N$. For example the
genetic algorithm presented in Section~\ref{sec:lifecycle-disc} always returns a
lifecycle model that is a sound, free-choice workflow net.

 By $\mathit{enabled}(t_o)$ we denote the 
composed expression (of ``pre-conditions") of the guard sentry for a stage/transition $t_o$. 
It is true if all necessary tokens for firing $t_o$ are available and the needed branch
conditions are true. As a guard sentry it will ensure that, as soon as this is the case,
the stage will open.

By $N$ being a
free-choice net, the enabling condition of transition $t$ is essentially a
positive boolean formula of conjunctions and disjunctions over the
\emph{visible} predecessors of $t$: for $t$ to be enabled, there has to be a
token in each pre-place $p \in \mathit{pre}(t)$ of $t$, and a token in $p$ is
produced by \emph{one} pre-transition $s \in \mathit{pre}(p)$ of $p$.
In addition, the occurrence of
transition $t$ itself can be subject to a condition represented by a variable $\mathit{cond}_t$
in the information model of the artifact. We assume that in general $\mathit{cond}_t$
can change its value during the lifecycle of the instance. 

Also, $init$ denotes the specific expression used to represent the event of the creation of an artifact instance, e.g. ``onCreate()". 

We can then define the predicate $\mathit{enabled}(t)$, assuming an ``origin" $t_o$, by a recursive definition as follows:
$$
\mathit{enabled}(t) = (\bigwedge_{p \in \mathit{pre}(t)} \mathit{markable}(p,t)) \wedge \mathit{cond}_t
$$
stating that transition $t$ is enabled iff its guarding condition is
satisfied and each pre-place $p$ can be marked. The predicate $\mathit{markable}(p,t)$ is defined as:
$$
\mathit{markable}(p,t) = \left\{
    \begin{array}{ll}
        \mathit{init}, & p\ \mbox{initially marked}\\
        \bigvee_{r \in \mathit{pre}(p)} \mathit{occurred}(r,t), & \mbox{otherwise}
    \end{array}
    \right.
$$
stating that place $p$ is either initially marked or can become marked by an
occurrence of any of its directly preceding transitions. A directly preceding
transition $r$ is either visible or invisible. 
The predicate $\mathit{occurred}(r,t)$ is then defined as follows:
$$
\mathit{occurred}(r,t) = \left\{
    \begin{array}{ll}
        \mathit{tokenAvailable}(r,t_o), & r\ \mbox{visible}\\
        \mathit{enabled}(r), & \mbox{otherwise}
    \end{array}
    \right.
$$
Here, as mentioned earlier, $\mathit{tokenAvailable}(t_p,t_o)$ is the specific expression that will be added to the sentry condition 
for each relevant visible transition $t_p$ with respect to the ``origin" $t_o$. 
The condition has to be precise enough to resolve any non-determinism between the guards of the subsequent stages connected to this sentry.
The format of these conditions will be discussed in the next section.

The recursion in the above definition stops either when reaching an initially marked place or when
reaching all visible predecessors. Thus, the predicate is only well-defined
if the net $N$ contains no cycle of invisible transitions.

The expression for $\mathit{enabled}(t_o)$ can be represented in a tree structure in a straightforward way.
The internal nodes of the tree represent logical operators (``and" or ``or") which are applied on
their child branches. The leaves represent either transitions that need to fire (which will be represented in the guard sentry
by an expression $\mathit{tokenAvailable}(t_p,t_o)$ for the specific transition $t_p$ in the leaf)
or decision point conditions that need to be true in order for
the ``origin" transition $t_o$ to be able to fire. 
In the following we use the words tree and expression interchangeably since, 
in this context, they represent the same information.

\begin{figure}[t]\centering
    \includegraphics[width=.28\textheight]{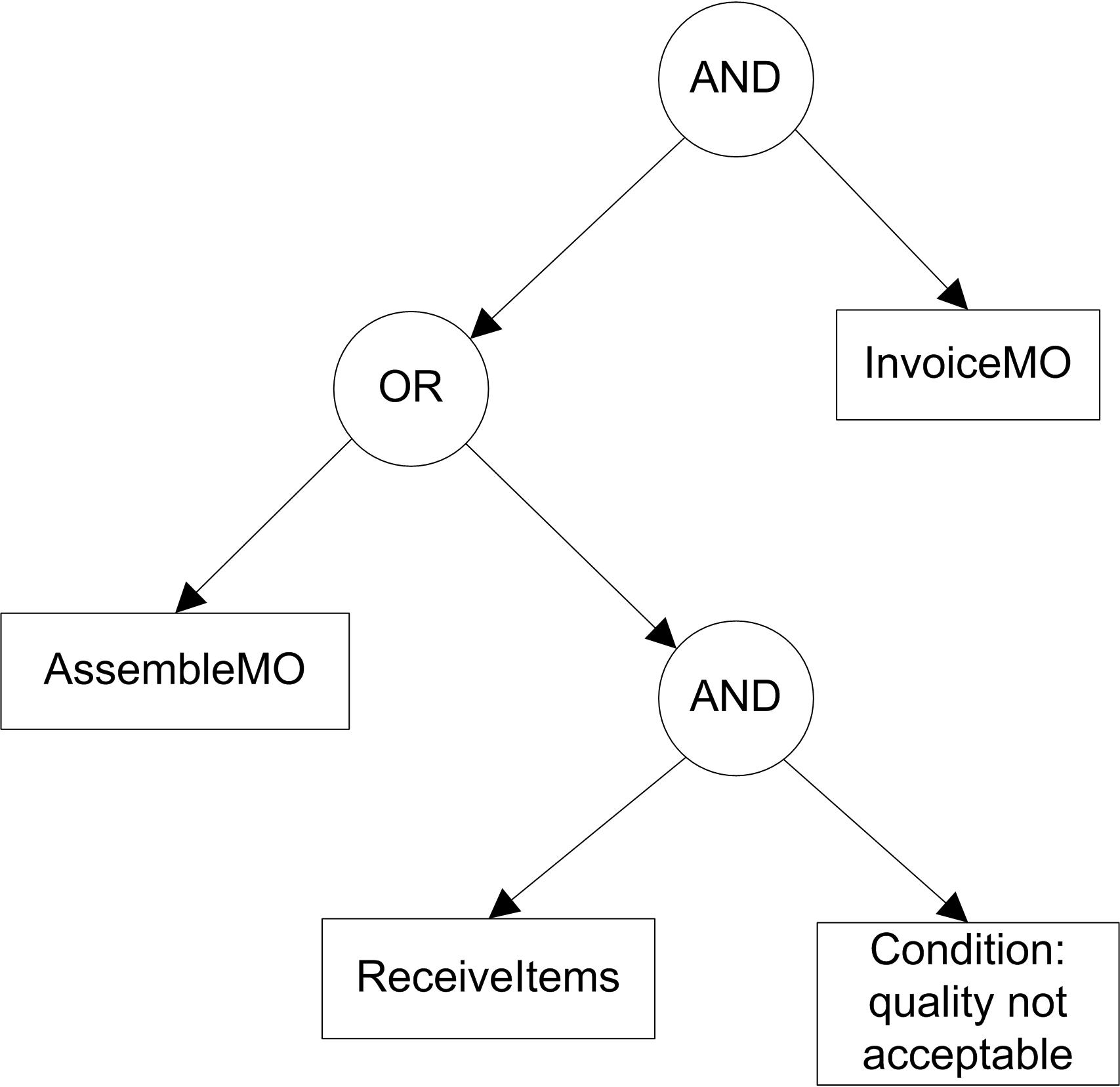}
    \caption{An example of an expression tree which will be used to generate the guard(s) for stage  $\fni{CompleteMO}$.}
    \label{fig:expression-tree}
\end{figure}

An example of such a tree is given in Fig.~\ref{fig:expression-tree} constructed for the transition
 $\fni{CompleteMO}$. Looking at the model in Fig.~\ref{fig:ordertocash} 
we can see that  $\fni{CompleteMO}$ can only fire if there is a token in each of its pre-places.
One of these tokens is generated by firing the transition   $\fni{InvoiceMO}$. 
The other token can arrive from two possible transitions -  $\fni{AssembleMO}$ or
the invisible transition represented as a black rectangle. We traverse back from 
the invisible transition and find out that it can only fire if the transition 
 $\fni{ReceiveItems}$ fires and the condition associated with the connecting arc is true 
(the received items have sufficient quality). This analysis results in the tree in
Fig.~\ref{fig:expression-tree}. The leaves of the tree are named by the corresponding transition 
or condition and, in fact, represent the specific expression for that
 transition/condition. However we delay the exact formulation of the expressions 
until the tree is built and analyzed, as will be described in the next section.

As mentioned earlier, an intermediate step of the algorithms decomposes $\mathit{enabled}(t_o)$ 
into several expressions which  are then used to generate separate guards of the stage.
Since $\mathit{enabled}(t_o)$ is a logical formula, we can convert it into Disjunctive Normal Form (DNF)
and assign each conjunction to a separate guard sentry. 

\begin{figure}[t]\centering
    \includegraphics[width=.38\textheight]{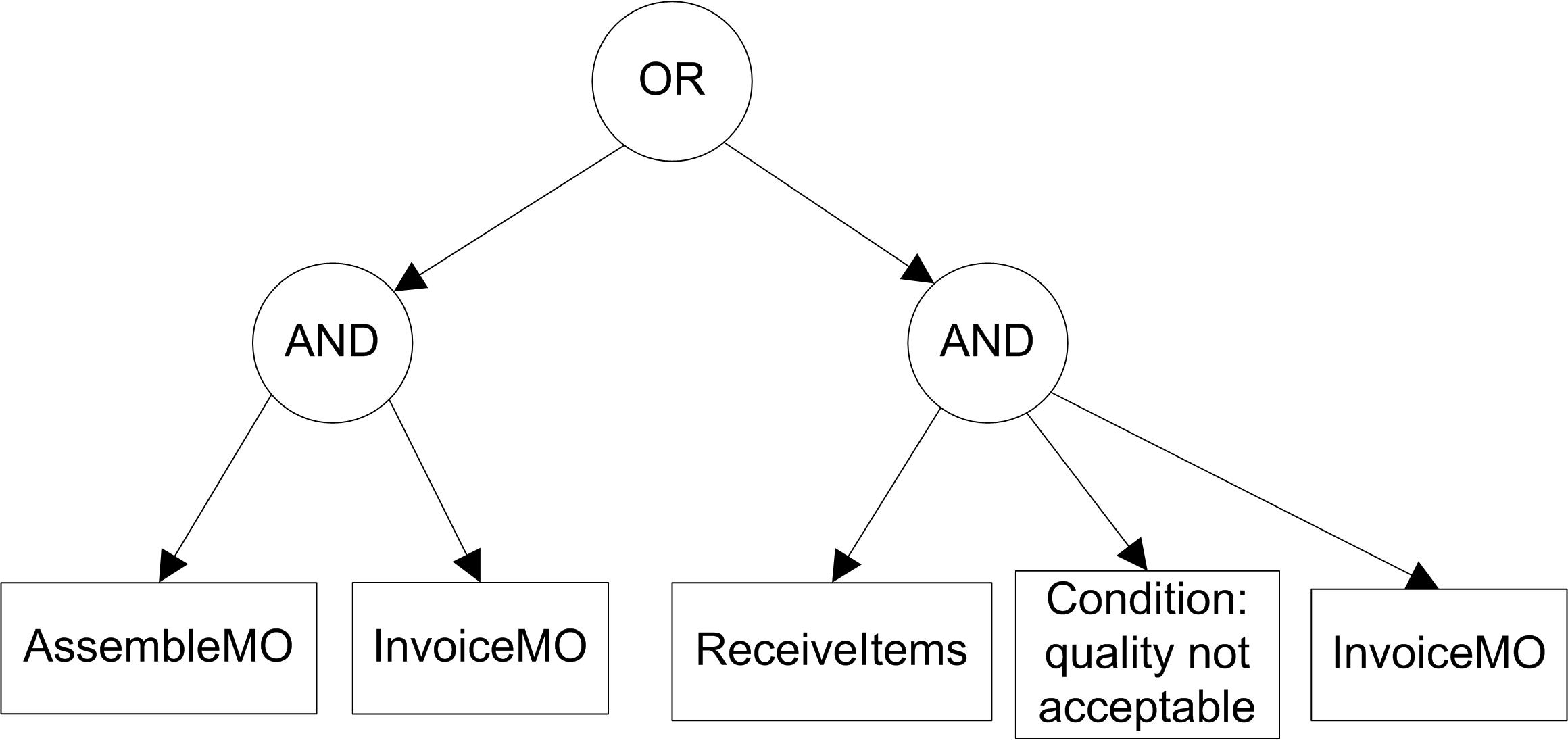}
    \caption{The expression tree for stage  $\fni{CompleteMO}$ in DNF.}
    \label{fig:expression-tree-dnf}
\end{figure}

After converting the example tree from Fig.~\ref{fig:expression-tree} into DNF, we now have the tree in Figure~\ref{fig:expression-tree-dnf}. 
Each child of the root node will generate one separate guard - here we have
two guards. Intuitively the first guard tells us that the stage will open if the items were assembled and invoice received. Similarly, the second
guard tells us that the stage will open if the items and invoice were received but the quality was insufficient to perform the assembling step.

As a final step, the $\mathit{tokenAvailable}(t_p,t_o)$ for the leaves of the tree are assigned as discussed in the next section.
\subsection{Formats for Pre-condition Expressions}
\label{sec:expressions}
In this section we look into the expressions $\mathit{tokenAvailable}(t_p,t_o)$ in more details and define
their format.
The assignment of a specific format to the expressions is delayed until the end, after $\mathit{enabled}(t_o)$ is composed and, 
if needed, decomposed into separate sentries $\{\mathit{enabled}_1(t_o), ..., \mathit{enabled}_n(t_o)\}$
representing all alternative ways the stage can be open. 
Only then it can be decided which format each expression should take.
We consider two possible formats for the expression of $\mathit{tokenAvailable}(t_p,t_o)$ depending on the context
as discussed below. 

\subsubsection{A simple format for pre-condition expressions}
\label{sec:simple}

The most simple case is when $\mathit{enabled}_i(t_o)$ contains only one transition $t_p$ 
with its expression $\mathit{tokenAvailable}(t_p,t_o)$ and $init$
is not present in $\mathit{enabled}_i(t_o)$. Then $\mathit{tokenAvailable}(t_p,t_o)$ can be replaced 
by the event corresponding to the finished execution of the activity of $t_p$.
It can be expressed using the event of achieving the milestone of the stage of $t_p$ or, alternatively, the closing of that stage among other options.

For example, for $t_o = \fni{ReceiveSupplResponse}$ the expression tree contains only one leaf 
corresponding to the transition $t_p = \fni{ReceiveMO}$, i.e., the only way to enable $t_o$ is
by a token produced by $t_p$ and this token cannot be consumed by another transition. Then the expression
for $t_p$ and $t_o$ will be $\mathit{enabled}(t_o) = \mathit{tokenAvailable}(t_p,t_o) = $ ``on $\fni{ReceiveMOMilestoneAchieved()}$"
where $\fni{ReceiveMOMilestoneAchieved()}$ is the event generated by the system when the milestone of 
stage $\fni{ReceiveMO}$ is achieved, therefore the task associated with the stage was completed and
the stage is closed.

If this is not the case, i.e., multiple transitions are present, then a more complex version of the 
$\mathit{tokenAvailable}(t_p,t_o)$ expression needs to be included
since we cannot use more than one event in the sentry. 
This form of the expression is discussed in the following sub-section. 
 

\subsubsection{A complex format for pre-condition expressions}
\label{sec:complex}

A token produced by the visible transition $t_1$ is available for $t_2$ as
long as neither $t_2$, nor any transition that is alternative to $t_2$
consumed that token. In a free-choice Petri net, a transition $t$ is
alternative to $t_2$ if they both have a common preceding place $p$ that is itself
preceded by $t_1$. We say a node $x$ precedes a node $y$ in net $N$, written
$x \rightarrow y$ iff there is path from $x$ to $y$ along the arcs of $N$. If
this path only involves $\tau$-labeled transitions (and arbitrarily labeled
places), we write $x \stackrel{\tau}{\rightarrow} y$.

With this in mind, we now define the following set of transitions that succeed $t_p$
and are alternative to $t_o$, i.e., visible transitions that are connected to a place 
on the path from $t_p$ to $t_o$:

\[\mathit{Alt}(t_p,t_o)=\{t\mid \exists \text{ place }p: t_p\xrightarrow{\tau} p \xrightarrow{\tau} t_o \thickspace\wedge\thickspace p\xrightarrow{\tau} t  \}. \]

$\mathit{Alt}(t_p,t_o)$ are the set of transitions that ``compete" with $t_o$ for the token produced by $t_p$. Therefore, 
in order to represent the situation when a token is present in the pre-place of $t_o$ and the stage $t_o$ should be opened,
we need to consider whether any
of the ``alternative" transitions have occurred (and ``stolen" the token).
Note that, according to this definition, $t_o$ will also belong to the set.

Let us consider again $t_o = \fni{CompleteMO}$ and the expression tree in Fig. \ref{fig:expression-tree-dnf}. 
Here we cannot use the simple format of the expressions for each leaf since in each AND-subtree there is more than one
transition. 
For the right-most AND-subtree, let us consider the leaf
$t_p = \fni{ReceiveItems}$, $\mathit{Alt}(t_p,t_o) = \{\fni{CompleteMO, AssembleMO}\}$.
Looking at Fig. \ref{fig:ordertocash}, we can see that the transition $\fni{AssembleMO}$ is indeed 
an ``alternative" to the invisible transition in the path from $t_p$ to $t_o$ in the sense that it can ``steal" the token produced by
the transition $\fni{ReceiveItems}$ in the connecting place. 

As a second building block we define the expression $\mathit{executedAfter}(t_p,t_s)$ 
which is true when there is a new execution of $t_p$ 
which occurs after the last execution of $t_s$ (meaning that it is relevant for
triggering the opening of the stage of $t_o$) and false otherwise. In terms of the Petri net 
it will be true when $t_s$ has produced a token that can potentially enable $t_o$, if it does not
get ``stolen" by another transition in the $\mathit{Alt}(t_p,t_o)$ set. 

How $\mathit{executedAfter}(t_p,t_s)$ will be expressed in the specific implementation can vary. 
Here we show how this can be done using the state of a milestone
(achieved or not) and the time a milestone was last toggled. 
By $m_p.hasBeenAchieved$ we denote a Boolean variable in the information model of the artifact which is true if the milestone $m_p$
of stage $t_p$ is in state ``achieved" and false otherwise. For every milestone $m_p$ present in a stage of the artifact there is also
a variable in the information model $m_p.lastToggled$ which gives the latest time stamp when $m_p$ was achieved or invalidated
(i.e. toggled its state).

Therefore the expression looks as follows:
\[\mathit{executedAfter}(t_p,t_s)= m_p.hasBeenAchieved  \thickspace\wedge \thickspace (m_p.lastToggled > m_s.lastToggled).  \]
In other words, the milestone $m_p$ of $t_p$ is achieved and it was last toggled after the milestone $m_s$ of $t_s$. Here we rely on the fact that the milestone of a stage will be
invalidated as soon as the stage is reopened. This is ensured by including an invalidating sentry for each milestone.

If for all members of
$\mathit{Alt}(t_p,t_o)$ $\mathit{executedAfter}(t_p,t_s)$ is true then the token is still
available to enable $t_o$. We express that as follows:
\[\mathit{tokenAvailable}(t_p,t_o) = \bigwedge_{t_s \in \mathit{Alt(t_p,t_o)}} \mathit{executedAfter}(t_p,t_s). \]

Of course, as discussed in the previous section,
other tokens produced in different branches of the Petri net might also be needed for enabling $t_o$
as well as the relevant branch conditions need to be true in order for $t_o$ to fire.

As an example, consider transitions $t_o = \fni{CompleteMO}$, $t_p = \fni{ReceiveItems}$
and $t_s = \fni{AssembleMO}$,
\begin{align*}
\mathit{tokenAvailable}(t_p,t_o) =& \thickspace \mathit{executedAfter}(t_p,t_s) \thickspace  \wedge \thickspace \mathit{executedAfter}(t_p,t_o)= \\
 =& \thickspace m_p.hasBeenAchieved  \thickspace\wedge \thickspace ( m_p.lastToggled > m_s.lastToggled ) \thickspace \wedge \\
 & \wedge (m_p.lastToggled > m_o.lastToggled), 
 \end{align*}
in other words, $\fni{ReceiveItems}$ was executed after the last execution of
 $\fni{AssembleMO}$ and after the last execution of  $\fni{CompleteMO}$,
 i.e., the token in the connecting place has not been consumed yet.
 
 This format is more general than the simple format and can be used in all cases. However it is less intuitive
 and therefore should be replaced by the simple format wherever possible. A third format of intermediate
 complexity can also be used which requires additional analysis of the Petri net 
for example for discovering the presence and location of cycles.
 This format is not presented here in order to simplify the discussion.

Using the proposed approach for translating the Petri net in Fig.~\ref{fig:ordertocash} to GSM, 
we generate the model in Fig.~\ref{fig:ordertocash-GSM} with guards as listed in Fig.~\ref{fig:ordertocash-guards} 

\begin{figure}\centering
{\scriptsize\sffamily
\begin{tabular}{|l|l|} \hline
\bf{Stage}  	&	\bf{Guard}       \\ \hline
$\fni{CreateMO}$      &	onCreate()   \\\hline
$\fni{ReceiveMO}$      &	on CreateMOMilestoneAchieved() 	 \\ \hline
$\fni{ReceiveSupplResponse}$      &	on ReceiveMOMilestoneAchieved()	 \\\hline
$\fni{ReassignSupplier}$      &	on ReceiveSupplResponseMilestoneAchieved() if answer = reject	 \\ \hline
$\fni{InvoiceMO}$      &	on ReceiveSupplResponseMilestoneAchieved() if answer = accept	 \\\hline
$\fni{ReceiveItems}$      &	on ReceiveSupplResponseMilestoneAchieved() if answer = accept  \\ \hline
$\fni{AssembleMO}$        &	on ReceiveItemsMilestoneAchieved() if quality = acceptable  \\ \hline
$\fni{CompleteMO}$      &  if InvoiceMOMilestone.hasBeenAchieved = true \\
                         &  and AssembleMOMilestone.hasBeenAchieved = true \\
						&  and InvoiceMOMilestone.lastToggled $>$ CompleteMOMilestone.lastToggled \\
                         &  and AssembleMOMilestone.lastToggled $>$ CompleteMOMilestone.lastToggled  \\\hline
$\fni{CompleteMO}$      &	if InvoiceMOMilestone.hasBeenAchieved = true \\
                         & and ReceiveItemsMilestone.hasBeenAchieved = true \\
						&  and InvoiceMOMilestone.lastToggled $>$ CompleteMOMilestone.lastToggled \\
                        &  and ReceiveItemsMilestone.lastToggled $>$ CompleteMOMilestone.lastToggled \\
                         &  and ReceiveItemsMilestone.lastToggled $>$ AssembleMOMilestone.lastToggled \\
                         & and quality = notacceptable \\\hline
$\fni{CloseMO}$      &	on CompleteMOMilestoneAchieved() \\\hline
$\fni{CloseMO}$      &	on ReassignSupplierMilestoneAchieved()  \\\hline
\end{tabular}

} \caption{The guards generated for the GSM model in Fig.~\ref{fig:ordertocash-GSM}}
\label{fig:ordertocash-guards}
\end{figure}

%
%
\section{Conclusions}
\label{sec:conclusions}
This paper presented a chain of methods for discovering artifact lifecycles which decomposes the problem
in such a way that a wide range of existing process discovery methods and tools can be reused. The proposed
tool chain allows for great flexibility in choosing the process mining methods best suited for the
specific business process. 

The presentation concentrates mostly on the artifact lifecycles. 
Additionally, the artifact information model can be built from the logs by extracting the data
attributes for each event type of the artifact. Existing tools such as~\cite{dec-miner} can be used
to mine data-dependent conditions for the guards based on the discovered information model.

The methods in this paper generate a flat model where no hierarchy of stages is used. Future
work will also consider methods for stage aggregation. One possible solution is to use
existing algorithms for process abstraction (e.g.~\cite{hierarchy1, hierarchy2}) for business process models and translate the
discovered process hierarchy to GSM stage hierarchy. For example the Refined Process Structure 
Tree~\cite{rpst} can be a first step to discovering such a hierarchy.

Future work will also develop
methods that allow to discover the interactions between artifacts and thus multi-artifact
GSM models can be generated such that instances of different artifacts can synchronize their lifecycles
in various ways.

\section*{Acknowledgment}
The research leading to these results has received funding from the European
Community's Seventh Framework Programme FP7/2007-2013 under grant agreement no 257593
(ACSI).

%
%

%

\begin{thebibliography}{5}

\bibitem{conformance2} Adriansyah, A., van Dongen, B.F., van der Aalst, W.M.P.: Towards Robust Conformance
Checking. In: Muehlen, M.z., Su, J. (eds.) Business Process Management Workshops.
LNBIP, vol. 66, 122–-133. Springer, Heidelberg (2011).

\bibitem{clustering} Bose, R.J.C., van der Aalst, W.: Trace Alignment in Process Mining: Opportunities for Process Diagnostics. 
In Business Process Management, vol. 6336, Springer Berlin / Heidelberg, 227-–242 (2010).

\bibitem{hierarchy1} Bose, R.P.J.C., Verbeek H.M.W., van der Aalst, W.M.P.:
Discovering Hierarchical Process Models using ProM. In:
Proc. of the CAiSE Forum 2011, London, UK (CEUR Workshop Proceedings, Vol. 734, 33-40) 

\bibitem{tree-miner} Buijs, J., van Dongen, B., van der Aalst, W.: A genetic algorithm for discovering process
trees. In: Proceedings of the 2012 IEEE World Congress on Computational Intelligence (2012).

\bibitem{artifacts2} Cohn, D., Hull, R.: Business artifacts: A data-centric
approach to modeling business operations and processes. IEEE Data Eng. Bull., 32, 3-–9 (2009).

\bibitem{CookWolf95}  Cook, J. E., Wolf, A. L.: Automating Process Discovery through Event-Data Analysis. 
In: Proceedings of the 17th international conference on Software engineering, New York, NY, USA, 73–-82 (1995).

\bibitem{ga1} De Medeiros, A.,  Weijters, A., van der Aalst, W.: Genetic Process Mining: an Experimental Evaluation. 
Data Mining and Knowledge Discovery, vol. 14, no. 2, 245-–304 (2007).

\bibitem{FahlandLDA11} Fahland, D., De Leoni,  M., Van Dongen, B. F., van der Aalst, W. M. P.: Many-to-many:
Some observations on interactions in artifact choreographies. In: Proc. of 3rd Central-European Workshop on Services and their Composition(ZEUS), 9--15. CEUR-WS.org (2011).

\bibitem{simplification} Fahland, D., van det Aalst, W.M.P.: Simplifying Mined Process Models: An Approach Based on Unfoldings. 
In: Proc. of BPM, 362--378 (2011).

\bibitem{repair} Fahland, D., van der Aalst, W.M.P.: Repairing process models to reflect reality. In: Business Process Management 2012, volume 7481 of Lecture Notes in Computer Science, 
229–-245. Springer (2012). 

\bibitem{Feldman-86} Feldman, P., Miller, D.: Entity model clustering: Structuring a data model by abstraction.
The Computer Journal, 29(4), 348--360 (1986).

\bibitem{fuzzy2} G{\"u}nther, C., van der Aalst, W.: Fuzzy Mining – Adaptive Process Simplification Based on Multi-perspective Metrics. 
In: Business Process Management, vol. 4714, Springer Berlin / Heidelberg, 328-–343, (2007).

\bibitem{hierarchy2} G{\"u}nther, C., van der Aalst, W.: Mining Activity Clusters from Low-Level Event Logs, 
BETA Working Paper Series, WP 165, Eindhoven University of Technology, Eindhoven (2006). 


\bibitem{gsm2}  Hull, R. et al. Introducing the guard-stage-milestone approach for specifying business
entity lifecycles. In: Proc. of 7th Intl. Workshop on Web Services and Formal Methods (WS-FM 2010), LNCS 6551. Springer-Verlag (2010).

\bibitem{gsm1} Hull, R. et al: Business Artifacts with Guard-Stage-Milestone Lifecycles: Managing
Artifact Interactions with Conditions and Events. In: DEBS 2011, 51--62 (2011).

\bibitem{rpst} Johnson, R., Pearson, D., Pingali, K.: The Program Structure Tree: Computing Control Regions in Linear Time.
In: Proc. of the ACM SIGPLAN 1994 conference on Programming language design and implementation, 
171--185, ACM (1994). 

\bibitem{declarative2} Maggi, F., Mooij, A., van der Aalst, W.: User-guided discovery of declarative process models. 
In: IEEE Symposium on Computational Intelligence and Data Mining, 192--199 (2011).

\bibitem{petri-nets} Murata, T.: Petri nets: Properties, Analysis and Applications. In: Proc. of the
IEEE, 541--580 (1989).

\bibitem{artifacts1} Nigam, A., Caswell, N.S.: Business artifacts: An
approach to operational specification. IBM Systems Journal, 42(3), 428-–445 (2003).

\bibitem{spm-to-bpel} Ouyang, C., Dumas, M., Breutel, S. ter Hofstede, A.H.M.: Translating Standard
Process Models to BPEL. In: CAiSE'06, 417--432 (2006).

\bibitem{declarative1} Pesic, M., van der Aalst, W.: A Declarative Approach for Flexible Business Processes Management. 
In: Business Process Management Workshops. Springer Berlin / Heidelberg, 169--180 (2006).

\bibitem{dec-miner} Rozinat, A., van der Aalst, W.M.P.: Decision Mining in ProM. In: S. Dustdar, J.L. Fiadeiro, and A. Sheth, editors, BPM 2006, LNCS 4102 , 420-–425. 
Springer-Verlag (2006). 

\bibitem{conformance1} Rozinat, A., Van der Aalst, W.M.P.: Conformance Checking of Processes Based on Monitoring
Real Behavior. Information Systems 33, 64–-95 (2008).

\bibitem{SilberschatzKS:2001:database} Silberschatz, A.,  Korth, H. F.,  Sudarshan, S.: Database System Concepts, 4th Edition.
McGraw-Hill Book Company (2001).

\bibitem{pd-survey} Tiwari, A.,  Turner, C.J.,  Majeed, B.: A Review of Business Process Mining: State-of-the-Art and Future Trends. 
Business Process Management Journal, vol. 14, no. 1, 5–-22 (2008).

\bibitem{pm-book} van der Aalst, W.M.P.: Process Mining: Discovery, Conformance and Enhancement
of Business Processes. Springer-Verlag, Berlin (2011).

\bibitem{AalstBEW:2001:proclets} van der Aalst, W., Barthelmess,  P.,  Ellis, C.,  Wainer, J.: Proclets: A Framework for
Lightweight Interacting Workflow Processes. Int. J. Cooperative Inf. Syst., 10(4), 443--481 (2001)

\bibitem{ga2} van der Aalst, W., De Medeiros, A., Weijters, A.: Genetic Process Mining. In Applications and Theory of Petri Nets 2005, 
vol. 3536, Springer Berlin / Heidelberg, p. 985 (2005).

\bibitem{alpha} van der Aalst, W.M.P., Weijters, A.J.M.M., Maruster, L.: Workflow Mining:
Discovering Process Models from Event Logs. IEEE Transactions on Knowledge and Data Engineering, 16(9), 1128--1142 (2004).

\bibitem{fuzzy1} van Dongen, B. F.: Process Mining: Fuzzy Clustering and Performance Visualization, In: BPM Workshops, 158-–169, (2009).

\bibitem{V_B_vD_vdA@BPM2010} Verbeek, H., Buijs, J. C., van Dongen, B. F., van der Aalst, W. M. P.: Prom: The process
mining toolkit. In: Proc. of BPM 2010 Demonstration
Track, CEUR Workshop Proceedings, vol. 615, 2010.

\bibitem{heuristic} Weijters, A.J.M.M., Van der Aalst, W.M.P.: Rediscovering Workflow Models
from Event-Based Data using Little Thumb. Integrated Computer-Aided Engineering, 10(2), 151--162 (2003).

\bibitem{causal-nets} Weijters, A.J.M.M., Ribeiro, J.T.S.: Flexible Heuristics Miner (FHM). 
In: Proceedings of the IEEE Symposium on Computational Intelligence and Data Mining, CIDM 2011, IEEE Symposium Series on Computational Intelligence 2011, 310--317 (2011).

\bibitem{ilp} Van der Werf, J.M.E.M., Van Dongen, B.F., Hurkens, C.A.J., Serebrenik, A.:
In: K.M. van Hee, R. Valk (Eds.), Applications and Theory of Petri Nets, 
LNCS vol. 5062, 368--387, Springer (2008).


\end{thebibliography}
\end{document}